\documentclass{aastex631}

\usepackage{url}
\usepackage{hyperref}
\usepackage{amsmath, amssymb}
\usepackage{natbib}
\usepackage{ulem}
\usepackage{bm}
\usepackage[utf8]{inputenc}
\usepackage{color, colortbl}
\usepackage{soul}
\usepackage{savesym}
\savesymbol{tablenum}
\usepackage{siunitx}
\restoresymbol{SIX}{tablenum}
\usepackage{todonotes}
\usepackage{gensymb}

\DeclareSIUnit\parsec{pc}
\DeclareSIUnit\mag{mag}
\DeclareSIUnit\h{\textit{h}}
\DeclareSIUnit\days{d}
\defcitealias{Carrick2015}{C15}
\defcitealias{LilowNusser2021ConstrainedFlow}{LN21}
\defcitealias{Riess2021ComprehensiveSH0ES}{R21}

\newcommand{\hubblevaluesimpleestimatorcentral }{ 70.3 } 
 
\newcommand{\hubblevaluesimpleestimatorcutcentral }{ 72.2 } 
\newcommand{\hubblevaluefixpvcovsncutcosmowitherr }{ 71.0_{ -4.6 }^{ +4.7 } } 
 
\newcommand{\hubblevaluefitpvcovsncutcosmowitherr }{ 71.5_{ -4.5 }^{ +4.6 } }

\newcommand{\hubblevaluefitpvcovsncutLilowcentral }{ 71.8 }

\newcommand{\hubblevaluefixpvcovzcutCarrickcentral }{ 77.0 }

\newcommand{\hubblevaluefiducialwitherr }{ 73.1_{ -2.3 }^{ +2.6 } }

\newcommand{\hubbleunit}{\kilo\meter\per\second\per\mega\parsec}

\newcommand{\vext}{V_\text{Ext.}}
\newcommand{\distance}{\chi}

\newcommand{\Einstein}{NASA Einstein Fellow}
\newcommand{\UCSC}{Department of Astronomy and Astrophysics, University of California, Santa Cruz, CA 95064}
\newcommand{\JHU}{William H. Miller III Department of Physics and Astronomy, The Johns Hopkins University, Baltimore, MD 21218.}
\newcommand{\STScI}{Space Telescope Science Institute, Baltimore, MD 21218}
\newcommand{\Duke}{Department of Physics, Duke University, Durham, North Carolina 27708}
\newcommand{\Texas}{Department of Physics and Astronomy, Texas A\&M University, College Station Texas 77843}
\newcommand{\Harvard}{Harvard-Smithsonian Center for Astrophysics, Cambridge, MA 02138}

\shortauthors{Kenworthy et al.}
\graphicspath{{./}{figures/}}

\begin{document}

\title{Measurements of the Hubble Constant with a Two Rung Distance Ladder: 

Two Out of Three Ain’t Bad}

\author[0000-0002-5153-5983]{W. D'Arcy Kenworthy}
\affiliation{\JHU}
\author[0000-0002-6124-1196]{Adam G. Riess}
\affiliation{\JHU}
\altaffiliation{\STScI}
\author[0000-0002-4934-5849]{Daniel Scolnic}
\affiliation{\Duke}
\author[0000-0001-9420-6525]{Wenlong Yuan}
\affiliation{\JHU}
\author[0000-0002-0961-4653]{José Luis Bernal}
\affiliation{\JHU}
\author[0000-0001-5201-8374]{Dillon Brout}
\altaffiliation{\Einstein}
\affiliation{\Harvard}
\author{Stefano Cassertano}
\affiliation{\STScI}
\author[0000-0002-6230-0151]{David O. Jones}
\altaffiliation{\Einstein}
\affiliation{\UCSC}
\author[0000-0002-1775-4859]{Lucas Macri}
\affiliation{\Texas}
\author[0000-0001-8596-4746]{Erik R. Peterson}
\affiliation{\Duke}

\correspondingauthor{W. D'Arcy Kenworthy}
\email{darcy@darcykenworthy.com}

\begin{abstract}

The three rung distance ladder{, which calibrates Type Ia supernovae through stellar distances linked to geometric measurements,} provides the highest precision {direct measurement} of the Hubble constant. In light of the Hubble tension, {it is important} to test the individual components of the distance ladder. For this purpose, we report a measurement of the Hubble constant from 35 extragalactic Cepheid hosts measured by the SH0ES team, using their distances and redshifts at $cz \leq \SI{3300}{\kilo\meter\per\second}$, instead of any, more distant Type Ia supernovae, to measure the Hubble flow. The Cepheid distances are calibrated geometrically in the Milky Way, NGC 4258, and the Large Magellanic Cloud. {Peculiar velocities are a significant source of systematic uncertainty at $z\sim 0.01$, and we present a formalism for both mitigating and quantifying their effects,} making use of external reconstructions of the density and velocity fields in the nearby universe. {We identify a significant source of uncertainty originating from different assumptions about the selection criteria of this sample, whether distance or redshift limited, as it was assembled over three decades.}  Modeling these assumptions yields central values ranging from $H_0=\hubblevaluefitpvcovsncutLilowcentral$ to $\hubblevaluefixpvcovzcutCarrickcentral$ \SI{}{\hubbleunit}. Combining the four best fitting selection models yields $H_0=\hubblevaluefiducialwitherr $ as a fiducial result, at $2.6\sigma$ tension with Planck.  While Type Ia supernovae are essential for a precise measurement of $H_0$, unknown systematics in these supernovae are unlikely to be the source of the Hubble tension.
\end{abstract}


\section{Introduction}

The current rate of cosmic expansion, {identified} as the Hubble constant $H_0$, is a fundamental cosmological parameter which sets the age and scale of the universe. In recent times, the  unexplained discrepancy between local measures of $H_0$ based on local distance indicators (e.g., \citealp{Riess2021ComprehensiveSH0ES}) and predictions based on measurements of the early universe (e.g., \citealp{PlanckCollaboration2018}) has attracted attention as a ``Hubble tension'' \citep{Valentino2021InSolutions}. Predictions based on the early universe appear robust, producing the same value at high precision when assuming the $\Lambda$CDM model with or without the use of CMB {anisotropy} data \citep{Addison2018}. Direct measurements of the Hubble constant at low redshift have been conducted using a variety of distance indicators, including water mega-masers \citep{Pesce2020MegamaserConstraints}, type II supernovae \citep{Jaeger2022HubbleSupernovae}, Surface-Brightness Fluctuations \citep{Blakeslee2021HubbleSBF}, and  Tully-Fisher relations \citep{Schomber2020BaryonicTullyFisher,Kourkchi2020CF4TullyFisher}, and modelling of gravitationally lensed quasars \citep{Wong2020H0LiCOWProbes,Birrer2020TDCOSMOivProfiles}. However, the most precise measurements in the literature have come from the use of a three-rung distance ladder using Type Ia supernovae (SNe\,Ia) \citep{Riess2021ComprehensiveSH0ES,Freedman2021MeasurementsPerspective}.

Since $\sim$ 2005, the SH0ES team \citep{Riess2005CepheidConstant,Riess2021ComprehensiveSH0ES} has been engaged in a project to refurbish the distance ladder measurement of $H_0$ using the WFC3 and ACS instruments on the Hubble Space Telescope. These measurements have been based on a three rung distance ladder formed by 1) geometric measurements of the distance to Cepheid variable stars, 2)  Cepheid distance measurements of SN\,Ia host galaxies, and 3) standardized SNe\,Ia magnitudes found in the Hubble flow between $0.023<z<0.15$. {This redshift range was chosen to reduce the dependence of the result on peculiar motions at low redshifts and cosmological effects at higher redshifts, while retaining sufficient statistics to constrain the Hubble flow.} Hallmarks of the SH0ES project include the use of the same instrument to measure all Cepheids across the ladder to nullify systematic uncertainties, the use of the NIR and three colors to mitigate dust and accounting for the dependence of Cepheid luminosity on locally measured abundances. This program also extended the range of Cepheid parallax measurements to a few kpc using spatial scanning of {\it HST}  \citep{Riess2014ParallaxTelescope,Casertano2016ParallaxMajoris}. {The SH0ES team has expanded the sample of cosmological SNe Ia calibrated by Cepheids from a few in the Hubble Key Project \citep{Freedman2001FinalConstant} with high quality data to 42 in 37 host galaxies in the most recent result (\citealp{Riess2021ComprehensiveSH0ES}; hereafter \citetalias{Riess2021ComprehensiveSH0ES}).}

In this work we present a two-rung distance ladder based on geometrically calibrated Cepheid distances {at $z<0.01$} as a measure of the Hubble flow. {In the conventional three-rung distance ladder, extragalactic distances measured by a geometrically-calibrated distance indicator (e.g. Cepheid or Mira variables, the tip of the red giant branch) co-located with one or more SNe\,Ia calibrate the SN\,Ia absolute luminosity. However the redshifts of these galaxies are not involved in the determination of $H_0$.} Instead, the Hubble flow is constrained by a separate sample of SNe\,Ia, because their bright absolute magnitude ($M_B \sim -19$) allows them to be detected well into the Hubble flow. By using objects at redshifts $cz \gtrapprox \SI{7000}{\kilo\meter\per\second}$, the fractional effect of galactic peculiar velocities ($\sim \SI{300}{\kilo\meter\per\second}$) {will be smaller than SN\,Ia uncertainties}. \citet{Wu2017} found that, given the \citet{Riess2016} SN Ia sample, (uncorrected) cosmic variance would affect the SH0ES $H_0$ measurement by only $\sim 0.5\%$. As SNe in calibrator and Hubble flow samples are largely observed by the same telescopes and discovered by similar surveys, {many} calibration and demographic biases are expected to largely cancel out from the differential measurement and are further reduced by recalibration using wide angle surveys \citep{Scolnic2015,Brownsberger2021PantheonZeropoints}. {Semi-independent crosschecks of the third rung of the distance-ladder include \citet{Burns2018CSPHubble}, \citet{Jones2022CosmologicalRAISIN}, and \citet{Dhawan2022UniformMethod}. However the astrophysics of Type Ia supernovae are debated, including the roles of progenitor age \citep{Jones2018,Rigault2020StrongRate,Rose2021HostDistances} and host galaxy dust \citep{Brout2020,Gonzalez2021EffectsScience,Thorp2021TestingDr1,Popovic2021Dust2Dust} in the diversity of observed populations.
SN-independent measures are thus} a crucial check of possible systematic biases in the SH0ES result.

In this work we remove the third rung, measuring $H_0$ by directly comparing Cepheid distance measurements to their host galaxy redshifts. { Doing so will give us a measurement of $H_0$ independent of any potential Type Ia supernova systematics. However by removing the third rung, we are forced to reduce the median redshift of the Hubble flow sample by an order of magnitude from $z \sim 0.04$ to $z \sim 0.006$ and the maximum redshift from $z=0.15$ to $z=0.011$.} At such low redshifts, peculiar velocities (PVs) are {expected to have a magnitude $\sim 20\%$ of the observed redshifts}. Further, the volume probed by these Cepheid hosts is not large compared to cosmological structures on the scale of baryon acoustic oscillations ($\sim \SI{140}{\mega\parsec}$), and as a result the effects of large-scale structure or ``cosmic variance'' on the measurement must be addressed. The reduction in median redshift by a factor of six can be naively expected to increase the effect of PVs on the measurement by at least the same amount, corresponding to an expected uncertainty in $H_0$ greater than $3\%$.

Cosmological measurements using Type Ia supernovae as standard candles have made use of peculiar velocity corrections based on models of the density field in the nearby universe to mitigate systematic PV effects from structure on scales greater than $\sim \SI{10}{\mega\parsec\per\h}$, { where $h=H_0 / \SI{100}{\hubbleunit}$}. Redshifts used in the cosmological analysis of \citet{Brout2022PantheonPlus} and \citet{Riess2021ComprehensiveSH0ES} also made use of ``group'' corrections (the replacement of an SN Ia host redshift with a mean redshift of associated galaxies) which attempt to remove noise on small scales from virial motions of galaxies within groups or clusters. \citet{Peterson2021PVCorrections} conclusively demonstrated at high significance that the dispersion of the SN\,Ia Hubble diagram is reduced with the application of both kinds of PV corrections. In particular two density field reconstructions \citep{Carrick2015,LilowNusser2021ConstrainedFlow} performed excellently on these tests, reducing the scatter of the Hubble diagram by $\sim 10\%$ in a sample of 584 SNe at $z<0.08$\footnote{The Peterson sample included the same galaxies measured by SH0ES Cepheids in their sample of SNe\,Ia host galaxies}. Both reconstructions are based on galaxy counts {and luminosities} from redshift surveys of the nearby universe, covering a cubic volume centered on the Milky Way with \SI{400}{\mega\parsec\per\h} sides.

In order to make a precise constraint on $H_0$ from  the Cepheid distances and Pantheon+ redshifts, it will be essential to make use of these reconstructions to extract as much information about the structure of the nearby universe as is possible. {To ensure our constraint is accurate we must also} quantify any remaining systematic uncertainties in the corrected redshifts, including the covariance between host galaxies. Finally, as the velocity noise relative to the distance noise increases at low redshifts, selection effects become increasingly important; the well known volumetric or Malmquist bias increases as the square of the relative error. At distances where PV effects make up $\sim 20\%$ of the signal of the Hubble flow, it will be necessary to evaluate the selection of the SH0ES Cepheid sample as fair markers of the Hubble flow and ensure that any associated uncertainties are included in our error estimates.

In this work, we make use of distances to Cepheid-hosting galaxies measured by SH0ES and Pantheon+ redshifts, in conjunction with two density field reconstructions of the nearby universe to measure the Hubble constant. In Section \ref{sec:Data} we discuss the data we make use of, including Cepheid data, redshifts, and PV reconstructions. In Section \ref{sec:method} we present our analysis methodology, which uses a hierarchical Bayesian model to address uncertainties and biases associated with a low-redshift measurement. This section includes estimates of the correlations in predicted redshifts based on modelling and the limitations of the reconstructions, as well as modelling of the SH0ES host selection. We discuss the implementation of this model in Section \ref{sec:implementation} and provide validations of the methodology on the SH0ES data. We present estimates of $H_0$ from the two-rung distance ladder in Section \ref{sec:results}, and discuss our findings in Section \ref{sec:conclusions}.

\section{Data}
\label{sec:Data}

\subsection{Cepheid Distances}

We use the distance moduli to Cepheid host galaxies measured by the SH0ES team. These are based on HST observations collected over 20 years using the WFC3 and ACS instruments in three optical bands (F555W, F814W, and F350LP) along with near-infrared (NIR) measurements in the F160W band . Sources for the data are given in Table 1 and Figure 2 of \citet{Riess2021ComprehensiveSH0ES}. These data have been reanalyzed using an automated pipeline for Cepheid detection, photometry, and data reduction described in Yuan et al. (2022b, in prep). We make use of dereddened Wesenheit magnitudes, and standardize using luminosity-period and -metallicity relations described in Sections 2 and 3 of \citet{Riess2021ComprehensiveSH0ES} . The absolute luminosities for the Cepheid-derived distances is constrained by three independent geometric distance measures:
\begin{enumerate}
    \item Milky Way Cepheid parallaxes from  \textit{Gaia} {EDR3} \citep{Gaia2016TheGaiaMission,Gaia2021EDR3, Riess2021GaiaEDR3} and HST spatial scanning \citep{Riess2018b}.
    \item An angular distance measurement to the water megamaser system in NGC 4258. \citep{Reid2019}
    \item A detached eclipsing binary distance to the Large Magellanic Cloud. \citep{Pietrzynski2019LMCDistance}
\end{enumerate}

These three anchors yield independent and consistent estimates of the fiducial Cepheid absolute luminosity (see Figures 19 and 21 of \citetalias{Riess2021ComprehensiveSH0ES}), and the constraints are combined in the final analysis. Our final dataset includes measurements of the distances to 35 unique galaxies within the Hubble flow which form a coherently constructed and relatively complete sample (targeting all suitable SN Ia-Cepheid hosts at $z\leq0.01$ with SN Ia seen 1980-2021) whose selection can be modelled; we have excluded from the analysis two galaxies which were targeted by a different HST program (HST-GO Proposal \#16269) because it used a very different selection method, targeting more luminous galaxies at greater redshifts of $0.01 < z < 0.02$. These data will be presented and analyzed in upcoming work.

We use the reduced data products of these efforts, specifically the set of distance moduli and associated covariances (which we label $\boldsymbol{\mu }_\textrm{Ceph},\boldsymbol{\Sigma}^{(\mathrm{SH0ES})}_\mu$) produced by the SH0ES analysis of \citetalias{Riess2021ComprehensiveSH0ES}. This covariance matrix encapsulates the statistical error in each distance as well as the correlated effects of crowding, metallicity/period standardization, and the three geometric anchor distances. As discussed in Appendix \ref{app:cephsys}, we include additional methodological uncertainties based on the analysis variants of \citetalias{Riess2021ComprehensiveSH0ES} to compute a final covariance matrix $\boldsymbol{\Sigma}_\mu$.

{While there are other extragalactic Cepheid measurements in the literature, we restrict ourselves to the SH0ES sample in order to avoid additional systematics relating to their selection functions, calibration uncertainties, photometry methodology, crowding corrections, and other methodological differences.}

\subsection{Galaxy Redshifts \& Peculiar Velocities}

We use the measured heliocentric redshifts of the 35 Cepheid galaxies taken from the Pantheon+ sample \citep{Carr2021PantheonRedshifts}. These are corrected for small-scale galactic motions by ``group corrections'', as presented in \citet{Peterson2021PVCorrections}, based on the galaxy-group assignments of \citet{Tully15}.  These redshifts are then converted into the CMB reference frame. With additional corrections for large-scale structure, these redshifts were used for the cosmology analysis of \citet{Brout2022PantheonPlus}, which found that they were consistent with $\sim$ \SI{240}{\kilo\meter\per\second} of uncorrected peculiar velocity scatter. This value was found in a mixed sample of galaxies with or without group corrections. We denote these corrected values as $\boldsymbol{z_\text{CMB}}$. {The sky coordinates for the host galaxies are also taken from \citet{Carr2021PantheonRedshifts}. We convert these sky coordinates into unit vectors in galactic Cartesian coordinates which we label $\hat{n}$.}

To correct these redshifts for large-scale structure, we use two reconstructions of the density and peculiar velocity fields in the nearby universe based on data from redshift surveys. These attempt to estimate the underlying fields that produce observed redshift distributions within the observed volume. \citet{Peterson2021PVCorrections} found that the two published reconstructions that most accurately match the SN\,Ia data are presented by \citet[hereafter \citetalias{Carrick2015}]{Carrick2015} and \citet[hereafter \citetalias{LilowNusser2021ConstrainedFlow}]{LilowNusser2021ConstrainedFlow}. In both cases the published versions are smoothed with a Gaussian kernel (at a scale of $5\hbox{--}10\ \SI{}{\mega\parsec\per\h}$) to remove difficult-to-model nonlinear behavior and to ameliorate the effects of shot noise. For convenience of comparison, we apply additional smoothing to the \citetalias{Carrick2015} products such that both are smoothed at scale of \SI{7}{\mega\parsec\per\h}. We will model redshift uncertainties due to limitations of these reconstructions in the following section.

The reconstruction of \citetalias{LilowNusser2021ConstrainedFlow} is based on redshifts compiled by the 2MASS Redshift Survey (2MRS) (\citealp{Huchra20122MASSRelease}; additional data released in \citealp{Macri20192MASSAvoidance}), using a bias-corrected Weiner filter estimate of the density field. That of \citetalias{Carrick2015} is based on the 2M++ redshift compilation \citep{Lavaux20112M++catalogue}, and used a method in which PVs and densities were iteratively recalculated until a best fit solution was found. {As the 2M++ compilation included the 2MRS sample released \citet{Huchra20122MASSRelease}, there is substantial data shared between the two efforts.}

Other reconstructions of the local matter and velocity fields have been described in published work, including \citet{Graziani2019CF3Field} and \citet{Jasche2021BORG}. We have chosen to make use of \citetalias{Carrick2015} and \citetalias{LilowNusser2021ConstrainedFlow} primarily due to their performance on the SN\,Ia data and the public accessibility of their data and data products.

\section{Methodology }
\label{sec:method}
 {Because uncorrected PVs represent a dominant source of uncertainty, and }redshift or distance limits imposed by the sample selection may produce a bias, we proceed with a more extensive effort to model the {data} than is typically used for a three-rung distance ladder where similar effects are negligible.

We construct a hierarchical Bayesian model for the observed distance moduli of the Cepheid galaxies given the reconstructed density/velocity fields, observed redshifts, and observed positions on the sky. We model our data as determined by the Hubble constant and a set of nuisance parameters which includes the true comoving distance to each observed galaxy as a latent parameter. This formulation of the problem allows us to easily account for aspects of the analysis such as selection, while budgeting for many of the systematic uncertainties of our analysis. In Section \ref{subsec:likelihood} we present our account of the observed data in terms of the model. Our treatment of the \citetalias{Carrick2015} and \citetalias{LilowNusser2021ConstrainedFlow} reconstructions are presented in Sections \ref{subsec:pvpred} and \ref{subsec:pvcovariance}. We discuss the priors we use in Section \ref{subsec:priors}, and the modeling for systematic effects of selection in Section \ref{subsec:selectionpriors}. We summarize our model in Section \ref{subsec:modelsummary}.

\subsection{Data Likelihood}
\label{subsec:likelihood}

We first define the relation between the modeled parameters and the observed data. To track the relationships between distances, redshifts, and peculiar velocities, we define the parameter vector of latent comoving distances in Mpc $h^{-1}$ units for the 35 objects of our Cepheid galaxy sample $\boldsymbol{\distance}$.  {
In a universe described by the Friedman-Lemaître-Robertson-Walker metric, the cosmological redshift for the $i$th object is} 
\begin{equation}
    z_\text{Cosm.}(\boldsymbol{\distance}_i) = 100\cdot \boldsymbol{\distance}_i /c \times (1+ \frac{ q_0 + 1}{2} \cdot 100\cdot \boldsymbol{\distance}_i /c  + \frac{ j_0 + 2q_0 + 1}{6} (100\cdot \boldsymbol{\distance}_i  /c)^2  +O(\boldsymbol{\distance}_i ^3))
\end{equation}

and concurrently the distance modulus is 

\begin{equation}
    \mu(\boldsymbol{\distance}_i,H_0) = 5 \log_{10} \left( \frac{cz_\text{Cosm.}(\boldsymbol{\distance}_i) }{H_0} \times  (1+ \frac{1-q_0}{2}  z_\text{Cosm.}(\boldsymbol{\distance}_i) - \frac{1-q_0-3q_0^2 +j_0}{6} z_\text{Cosm.}(\boldsymbol{\distance}_i)^2 + O(\boldsymbol{\distance}_i ^3)) \right)+25
\end{equation}
with the two kinematic parameters fixed to $q_0=-0.55$ and $j_0=1$, and $H_0$ a free parameter. Our result is insensitive to changes in the kinematic quantities, as the effect of these corrections to the Hubble law is $\sim 0.5\%$ for the most distant objects in our sample, and much less for the majority. 

The likelihood of the observed distance moduli $\boldsymbol{\mu }_\textrm{Ceph}$ with uncertainties $\boldsymbol{\Sigma}_{\mu}$ is simply defined by the multivariate normal distribution

\begin{equation}
    \boldsymbol{\mu }_\textrm{Ceph}| \boldsymbol{\distance},\log H_0 \sim \mathcal{N}( \mu(\boldsymbol{\distance},H_0) , \boldsymbol{\Sigma}_{\mu}) \label{eq:mulikelihood}.
\end{equation}
\noindent This term of the model is the component of the likelihood in which $H_0$ appears.

The observed likelihood of the redshifts is more complicated. The redshift addition formula gives the observed redshift corrected to the CMB frame as $z_\text{CMB} +1 = (1+ z_\text{Pec.})\times (1+ z_\text{Cosm.})$. From this the likelihood of the observed vector of CMB frame, group-corrected redshifts $\boldsymbol{z_\text{CMB}}$ are given by the multivariate normal distribution

\begin{equation}
\frac{1+\boldsymbol{z_\text{CMB}}}{1+z_\text{Cosm.}(\boldsymbol{\distance})} | \boldsymbol{\distance},\boldsymbol{\theta}  \sim \mathcal{N}( 1+v_\text{Pred.}(\boldsymbol{\distance}, \boldsymbol{\theta})/c,\boldsymbol{\Sigma}_\text{Pec.}(\boldsymbol{\distance}; \boldsymbol{\theta})/c^2) \label{eq:zlikelihood}
\end{equation}
where the division on the left hand side is conducted elementwise. $v_\text{Pred.}(\boldsymbol{\distance},\boldsymbol{\theta})$ is the predicted peculiar velocity from the reconstruction used discussed in Section \ref{subsec:pvpred}, and $\boldsymbol{\Sigma}_\text{Pec.}(\boldsymbol{\distance}; \boldsymbol{\theta})$ is our estimated uncertainty in these predictions as presented in Section \ref{subsec:pvcovariance}. The parameters $\boldsymbol{\theta}$ that define the velocity predictions and uncertainties will be introduced in the next two sections.

These two likelihood definitions make up the constraints on the model parameters from the data. $H_0$ is measured as the relative scale between the Cepheid distances and the redshifts.

\subsubsection{Peculiar Velocity Predictions}
\label{subsec:pvpred}

The two peculiar velocity reconstructions are publicly distributed as data cubes which give the values of the  density contrast $\delta(\Vec{r})= (\rho(\Vec{r}) - \Bar{\rho} )/ \Bar{\rho} $ (where $\rho$ is the density of matter) and peculiar velocities $\Vec{v}_\mathrm{Pec.}(\Vec{r})$ at points on a densely sampled grid. These grids are given at a resolution higher than the smoothing scale. These can easily be interpolated to give functions for the density contrast and velocity as a function of position in comoving galactic coordinates $\delta_\text{Rec.}(\Vec{r}), \Vec{v}_\text{Rec.}(\Vec{r})$ respectively.

The two reconstructions also have both made measurements of the parameters that relate the density to observed velocities. {We quantify the effect of the stated uncertainties by marginalizing over the values given in the original work. The relevant parameters are:}
\begin{enumerate}
    \item The scale parameter $\beta$, the ratio of the {logarithmic growth rate $f(\Omega_M)$ of density perturbations} to the galaxy bias $b$. This relates predicted peculiar velocity to galaxy overdensities.
    \item The external bulk flow $\Vec{V}_\mathrm{Ext}$, a velocity vector added to every point within the reconstructed volume. This models any contribution to PVs sourced by perturbations on scales larger than the volume probed by the data.
\end{enumerate}

The external bulk flow is parameterized by $V_\text{Ext}$, the magnitude in units of \SI{}{\kilo\meter\per\second}, and $(l,b)$ the galactic longitude and latitude of its direction. Note that this dipole is distinct from the $\sim \SI{600}{\kilo\meter\per\second}$ motion of the Milky Way with respect to the cosmic microwave background, most of which is due to gravitational effects from the local volume (within \SI{200}{\mega\parsec\per\h}). We remove the fiducial values of the quantities $\{ \beta_\textrm{Fid.}, V_\text{Fid. Ext.}, l_\textrm{Fid.},b_\textrm{Fid.}\}$ from the calculated peculiar velocities to define the unscaled local velocities

\begin{equation}
    \Vec{u}_\text{Rec.}(\Vec{r}) =\frac{\Vec{v}_\text{Rec.}(\Vec{r})- V_\text{Fid. Ext.} \times  \hat{n}(l_\text{Fid.},b_\text{Fid.})}{\beta_\textrm{Fid.}}
\end{equation}
\noindent {where $\hat{n}(l,b)$ is the unit position vector as a function of angular coordinate.} We can then compute the predicted peculiar velocity for the $i$th object with unit position vector $\hat{n}_i$

\begin{equation}
    v_\text{Pred.}(\boldsymbol{\distance}_i,\beta,\vext,l,b) =  \hat{n}_i \cdot [ \beta  \Vec{u}_\text{Rec.}( \boldsymbol{\distance}_i \times \hat{n}_i) + \vext \times \hat{n}(l,b)  ) ].
\end{equation}

\subsubsection{Covariance of PV reconstructions}
\label{subsec:pvcovariance}

Previous work including \citet{Riess2016}, \citet{Scolnic2018}, \citet{Boruah2020CosmicConstraints} and others have analyzed SN\,Ia distances while accounting for PV uncertainties using an assumption of diagonal scatter in corrected redshifts, corresponding to an assumption that all correlations in the data have been removed. However, the smoothing of the density field on scales $\sim \SI{10}{\mega\parsec\per\h}$ removes any information at smaller distances. Any two galaxies closer together than this scale will have similar predictions for their density/PV, while their true velocities may include unmodelled nonlinear or virial noise. Further, they make use of a finite number of tracers (galaxies) of the underlying density. Lastly, the two reconstructions show some differences in predicted densities and velocities at the scale of $\sim \SI{60}{\kilo\meter\per\second}$, {with some differences apparent across the entire volume}. It is thus apparent that use of these reconstructions to remove predicted velocities will ameliorate but not totally remove correlations in the data, and a trustworthy estimate of the Hubble constant from data with median distance $\sim \SI{20}{\mega\parsec\per\h}$ will require an estimate of these correlations. Our goal then is to construct an appropriate covariance matrix $\boldsymbol{\Sigma}_\mathrm{Pec.}$ for the predicted peculiar velocities.  {These correlated uncertainties will ultimately contribute $\sim 2\%$ of the final error budget (see Table \ref{tab:modeldiffs}).}

To understand the uncertainties in the reconstructed density fields, we first need to understand {the scales and amplitudes} of the underlying cosmological structures {which these attempt to model}. If the density field under $\Lambda$CDM is approximated as Gaussian, the statistics of the density contrast $\delta$ are fully specified by the power spectrum $P(k)$. The power spectrum is defined by the relation

\begin{equation}
    <\Tilde{\delta}(\Vec{k}^*) \Tilde{\delta}(\Vec{k}) > = (2\pi)^3 P(|\Vec{k}|) \delta^3(\Vec{k}-\Vec{k}^*)
\end{equation}
where $\delta^3$ is a Dirac delta function and $\Tilde{\delta}(k)$ is the Fourier transform of the density contrast. Given the linear approximation that the divergence of the velocity field is proportional to the density and that the velocity is vorticity free, the covariance of the peculiar velocities along the line of sight at comoving distances $\distance_i$ and $\distance_j$ separated by angle $\theta$ is given by

\begin{align}
     <v_i v_j> &= \frac{dD}{d \tau}\Bigr|_{\distance_i} \frac{dD}{d \tau}\Bigr|_{\distance_j} \int dk / (2\pi^2)\ P_{\Lambda\mathrm{CDM}}(k) \times F(k {\distance}_{i},k {\distance}_{j}, \cos(\theta )) \\
    F(u,v,\cos(\theta)) &= \frac{\partial^2 \text{sinc}(\sqrt{x^2-2xy \cos(\theta) + y^2})}{\partial x \partial y}\bigg\rvert_{x=u,y=v}
\end{align}
where $\tau$ is conformal time, and $D$ is the {linear, scale-independent} ``growth factor'' that describes the 
{amplitude of density perturbations} as a function of cosmic time \citep{Davis2011,Huterer2015}. The term we write as $F(u,v,\cos(\theta))$ can be evaluated either as a series over spherical Bessell functions or constructed from elementary and trigonometric functions.  We evaluate the power spectrum using the CLASS code \citep{Blas2011CLASS} with nonlinear corrections from halofit \citep{Takahashi2012Halofit} using cosmological parameters from Planck \citep{PlanckCollaboration2018}. As $F(u,u,1)=1/3$, the predicted diagonal dispersion $<v_i^2>$ at $z=0$ is proportional to the integral $\int dk P(k)$. Evaluating $<v_i^2>$, we find a predicted scatter of $(\SI{380}{\kilo\meter\per\second})^2$ of peculiar velocity scatter in the local universe. {This quantity represents the uncertainty in uncorrected CMB-frame redshifts as a distance-indicator.}

\begin{figure}
    \centering
    \includegraphics{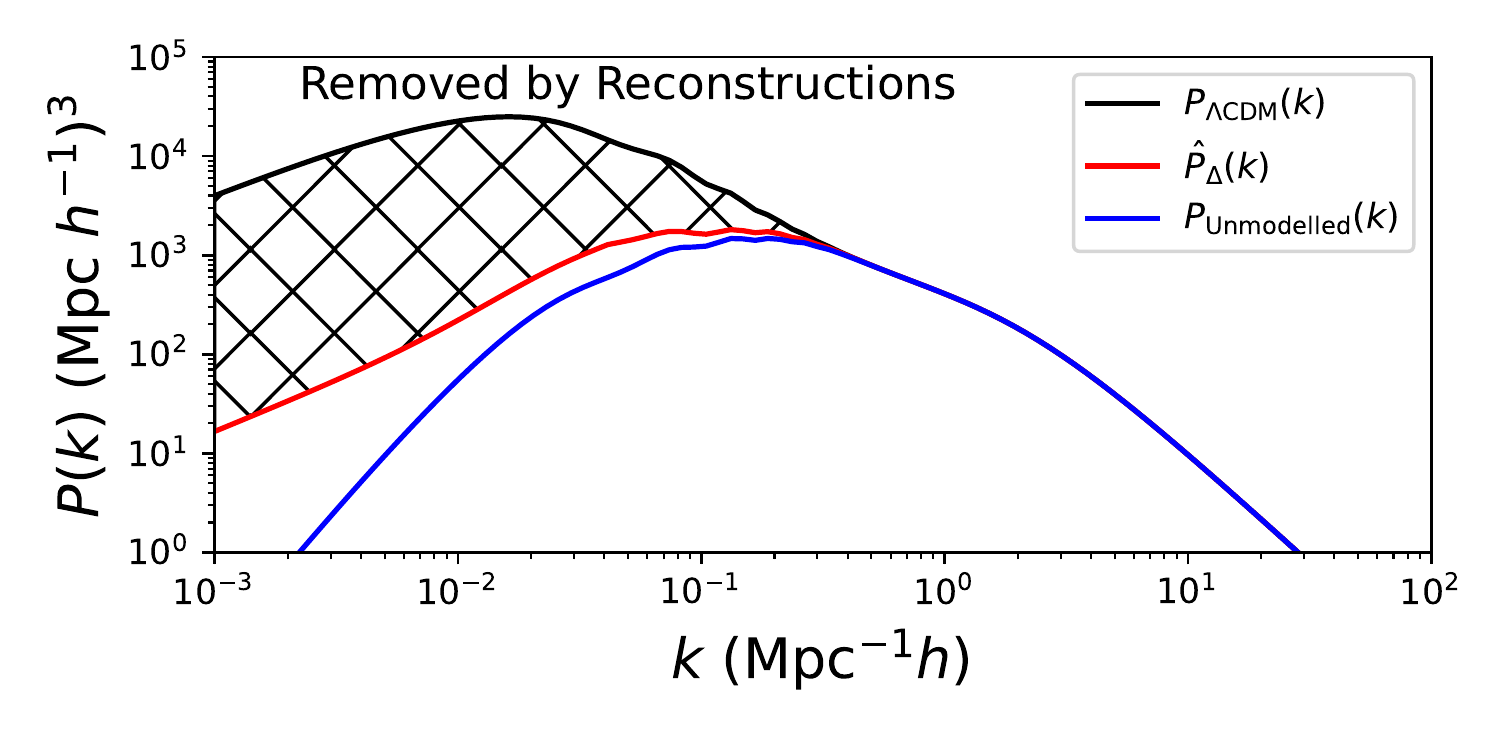}
    \caption{Comparison between the power spectrum of $\Lambda$CDM, as compared to the estimated power in the residuals after correction with a PV reconstruction. $P_\mathrm{Structure}(k)$ is the difference between red and blue lines. Amplitudes have been fixed to fiducial values. }
    \label{fig:powerestimated}
\end{figure}

{To budget for PV uncertainties in redshifts \textit{after} correction using a reconstruction, we derive two independent sources of remaining noise, which we label $P_\mathrm{Unmodelled}(k),P_\mathrm{Structure}(k)$. These quantities are derived and defined in Appendix \ref{app:pvcovariancederiv}}. $P_\mathrm{Unmodelled}(k)$ represents uncertainties  {resulting from the Gaussian smoothing applied in the reconstruction procedures}, and is predicted to contribute \SI{264}{\kilo\meter\per\second} of diagonal scatter. The size of this prediction is uncertain, as it is dependent on theory/simulation-based accounts of nonlinear gravitational effects. $P_\mathrm{Structure}(k)$ is empirically estimated from the differences between the results of \citetalias{Carrick2015} and \citetalias{LilowNusser2021ConstrainedFlow}, and represents our estimate of the systematic uncertainty in their modeling of large-scale structure. This component contributes \SI{66}{\kilo\meter\per\second} of diagonal scatter, which is present across larger distances on the sky. {The sum of these two components, which we label $\hat{P}_{\Delta}(k)$, is the total power remaining in the density (and thus velocity) field after correction. We compare $\hat{P}_{\Delta}(k)$ to the $\Lambda$CDM prediction in Figure \ref{fig:powerestimated}; the difference between the two represents the uncertainty removed from our measurement by the use of the galaxy catalog information. Accordingly, our fiducial error budget predicts use of the corrections will reduce the diagonal scatter of the sample from $\sim (\SI{380}{\kilo\meter\per\second})^2$ to $\sim (\SI{270}{\kilo\meter\per\second})^2$.}

To account for uncertainties in the size of each of the two effects discussed, we introduce two additional parameters $\sigma^\mathrm{(PV)}_\mathrm{Unmodelled}, \sigma^\mathrm{(PV)}_\mathrm{Structure}$ which scale the amplitude of the two components of the budgeted uncertainties:

\begin{equation}
    \hat{P}_{\Delta}(k; \sigma^\mathrm{(PV)}_\mathrm{Unmodelled}, \sigma^\mathrm{(PV)}_\mathrm{Structure})  =   \left({\sigma^\mathrm{(PV)}_\mathrm{Unmodelled}}\right)^2 P^{\mathrm{(Normalized)}}_\mathrm{Unmodelled}(k)+ \left({\sigma^\mathrm{(PV)}_\mathrm{Structure}}\right)^2 P^{\mathrm{(Normalized)}}_\mathrm{Structure}(k) \label{eq:corrpowerspec}
\end{equation}

\noindent where each component of the power spectrum has been normalized by the fiducial scatter ${<v_i^2>}$ so the value of each parameter is equal to the scatter contributed by each component. A value of $\sigma^\mathrm{(PV)}_\mathrm{Unmodelled}$ greater than the fiducial value of \SI{264}{\kilo\meter\per\second} would indicate large virial or nonlinear motions on scales too small to be detected by the reconstructions; a value of $\sigma^\mathrm{(PV)}_\mathrm{Structure}$ larger than the fiducial value of \SI{66}{\kilo\meter\per\second} would indicate that our estimates of the error in the reconstructed fields are understated.

{In two model variants (results shown in the first two rows of Table \ref{tab:results})} we attempt to fit the data without the use of external galaxy catalog information. In these cases, we instead use the relation

\begin{align}
    \hat{P}_{\Delta}(k; \sigma^\mathrm{(PV)}_\mathrm{Unmodelled}, \sigma^\mathrm{(PV)}_\mathrm{Structure})  &=   \left({\sigma^\mathrm{(PV)}_\mathrm{Unmodelled}}\right)^2 P^{\mathrm{(Normalized)}}_\mathrm{Unmodelled}(k)+ \left({\sigma^\mathrm{(PV)}_\mathrm{Structure}}\right)^2 P^{\mathrm{(Normalized)}}_{\Lambda\mathrm{CDM\ Structure} }(k) \label{eq:nocorrpowerspec}\\
\end{align}
{where $P_{\Lambda\mathrm{CDM\ Structure} }(k)$ is defined}
\begin{align}
    P_{\Lambda\mathrm{CDM\ Structure} }(k) &=  P_{\Lambda\mathrm{CDM} }(k) - P_\mathrm{Unmodelled}(k).
\end{align}

\noindent This reduces to the full $\Lambda$CDM power spectrum in the case when the amplitude parameters $\sigma^\mathrm{(PV)}_\mathrm{Unmodelled}, \sigma^\mathrm{(PV)}_\mathrm{Structure}$ are fixed to their fiducial values, but allows the measurement to marginalize over unknown nonlinearities and cosmology.

Our estimated peculiar velocity covariance is then
\begin{equation}
    \boldsymbol{\Sigma}_{Pec}( \boldsymbol{\distance}_{i}, \boldsymbol{\distance}_{j};\sigma^\mathrm{(PV)}_\mathrm{Unmodelled}, \sigma^\mathrm{(PV)}_\mathrm{Structure} ) = \frac{\partial D}{\partial \tau}^2 \int dk / (2\pi^2)\ \hat{P}_{\Delta}(k;\sigma^\mathrm{(PV)}_\mathrm{Unmodelled}, \sigma^\mathrm{(PV)}_\mathrm{Structure}) \times F(k \boldsymbol{\distance}_{i},k \boldsymbol{\distance}_{j}, {\hat{n}}^{(i)} \cdot {\hat{n}}^{(j)}).  \label{eq:PVcovariance}
\end{equation}
\noindent Three Cepheid hosts are identified as members of the Virgo cluster from NED, but have received different group assignments with distinct redshifts No other galaxies in our sample appear to be cluster members. To account for the difficulty of the group assignments in the dense environment with the cluster, we add an additional $(\SI{550}{\kilo\meter\per\second})^2$ of variance to the diagonal elements of the covariance matrix for these three objects, representing the virial noise of the complex \citep{Kashibadze2020StructureVirgo}. {An alternative approach could assign a single redshift for the cluster center to all three objects, but the width of the cluster is not negligible relative to the distance. Particularly given that identifying a unique estimate for the central redshift of an object with the size and complex structure of Virgo is nontrivial, we conclude our method is more robust. }

\begin{figure}
    \centering
    \includegraphics{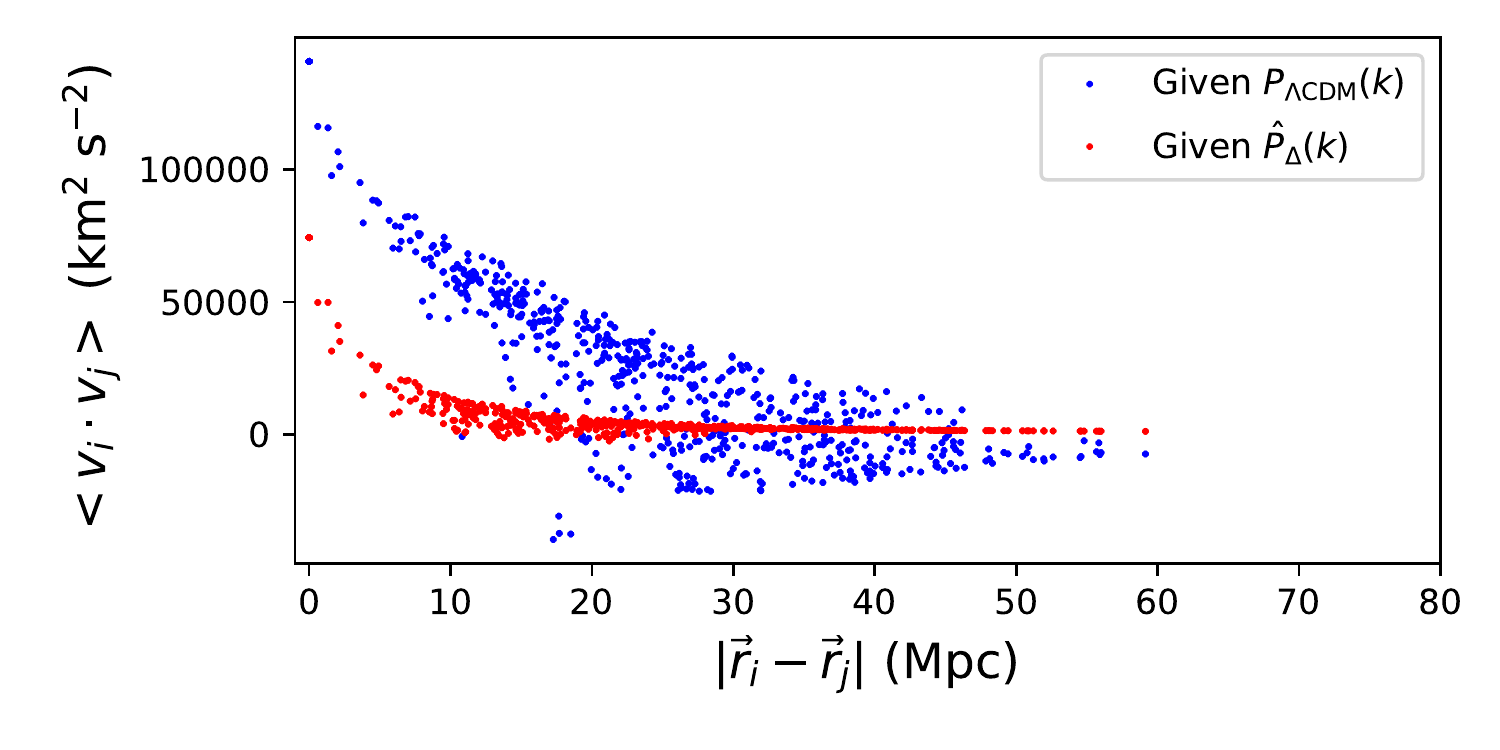}
    \caption{ Comparison of the velocity covariance given the power spectrum of $\Lambda$CDM in blue, and given the estimated power after corrections from PV reconstructions in red. Each point represents the covariance between a pair of Cepheid host galaxies. As we expect the overall variance of the sample is cut in half on the diagonal (zero separation), and the remaining correlations are present on much smaller scales. }
    \label{fig:vpeccovarianceseparation}
\end{figure}
We calculate the covariance matrix for our sample at fiducial distances using our estimated power spectrum $\hat{P}_{\Delta}(k) $ as well as $P_{ \Lambda\mathrm{CDM}}(k)$ at fixed amplitude. We 
{show the values of the covariance between each pair of objects compared to their physical separation in Figure \ref{fig:vpeccovarianceseparation}.  As can be seen,} applying the corrections to remove power at large scales drastically reduces the correlations internal to our sample. As the window function $F(u,v,\theta)$ is a complex function of the lines-of-sight of each of the two objects and their angle with respect to to each other, the points fall within an envelope as a function of physical separation. The upper bound is set by a particular geometry of the two objects with respect to the observer with small but nonzero angular separation, while the lower bound will be set by a geometry with the same physical separation but $180 \degree$ greater angular separation, which will have a similar scale of correlation with opposite sign. The sample of observed Cepheids has few such nearby pairs on opposite sides of the sky at small scales, thus the lower half of the envelope is not seen until larger scales.

\subsection{Priors \& Sample Definition}
\label{subsec:priors}
We use the following priors on the scalar parameters:
\begin{enumerate}
    \item $\log H_0 \sim \mathcal{N}(4,1)$
    \item $\beta \sim \mathcal{N}(\beta_\text{Fid.},\sigma_\beta)$
    \item $\vext \sim \mathcal{N}(V_\text{Fid. Ext.}, \sigma_V )$
    \item $l \sim \mathcal{N}(l_\text{Fid.}, \sigma_l )$
    \item $b \sim \mathcal{N}(b_\text{Fid.}, \sigma_b )$
    \item $\sigma^\mathrm{(PV)}_\mathrm{Structure} \sim \mathrm{InvGamma}(6.52,6.18 \sigma_\mathrm{Structure}^\mathrm{Fiducial}) $,
    $\sigma^\mathrm{(PV)}_\mathrm{Structure} \in (0,5 \sigma_\mathrm{Structure}^\mathrm{Fiducial}]$ 
    \item $\sigma^\mathrm{(PV)}_\mathrm{Unmodelled} \sim \mathrm{InvGamma}(3.58,3.25 \sigma_\mathrm{Unmodelled}^\mathrm{Fiducial}) $
\end{enumerate}

The fiducial quantities for the central value and uncertainties on the reconstruction parameters were taken directly from the relevant published work {(i.e. \citetalias{Carrick2015} or \citetalias{LilowNusser2021ConstrainedFlow})}. {For the PV amplitude parameters, we chose the inverse gamma distribution as a positive-definite but relatively uninformative distribution.} The scale and shape parameters of the distribution were chosen to restrict the model from accessing pathological regions of the posterior at extreme values of these parameters; they have medians centered on the fiducial values, with 1\% of probability mass at values greater than $3\times$ and $5\times$ their respective fiducial values . {{As previously given, the fiducial value of $\sigma^\mathrm{(PV)}_\mathrm{Unmodelled}$ is \SI{264}{\kilo\meter\per\second}. The fiducial value of  $\sigma^\mathrm{(PV)}_\mathrm{Structure}$ is dependent on whether we use reconstructions ($\SI{66}{\kilo\meter\per\second}$) or assume the full $\Lambda$CDM systematic (\SI{266}{\kilo\meter\per\second}).}} { We additionally put an upper bound on $\sigma_\mathrm{Structure}$ for computational reasons, as in some cases one MCMC chain could become stuck in the pathological region associated with high values of this parameter.}

\subsubsection{Selection \& Distance Priors}
\label{subsec:selectionpriors}

For nearby galaxies, the observed redshift as a proxy for distance is significantly biased {by volume effects}. Since the number density of galaxies per unit distance generally goes as the square of distance, a galaxy observed with a given redshift is likely to have scattered down from the larger volume element with higher true distance than would be naively inferred by the Hubble law. For a PV scatter of \SI{250}{\kilo\meter\per\second}, this bias is approximately given by 

\begin{equation}
    \frac{\Delta_z}{z} \approx \frac{ 10^{-6}}{z^2},
\end{equation}

\noindent derived from Monte Carlo simulations described below; for our sample, this estimated bias ranges from $\sim 100\%$ for the lowest redshift objects to $\sim 1\%$ at the edge of our sample.

In comparison, the Cepheid distances exhibit little statistical noise, resulting in much smaller bias. To quantify and account for the biases of our sample, it is necessary to understand the selection of our sample of Cepheid host galaxies. Specifically, it is necessary to check whether the fact a galaxy was included in our sample provides additional information about the galaxy's distance, independent from the measured values of the redshift and Cepheid distance modulus.

The selection goals of the SH0ES program were ultimately to target every star-forming galaxy at suitable inclination that has hosted a SN of cosmological quality reported before maximum light and observed digitally in the local volume at $d \lessapprox \SI{40}{\mega\parsec}$ or $z \lessapprox 0.011$. However there is a key ambiguity, in that the indicator used to evaluate this last criterion could be either the redshift of the galaxy or the SN Ia inferred distance. {The SH0ES program was conducted over $\sim$ 15 years and without any pre-allocation of observing time, so {\it HST} proposals were submitted on a cycle-to-cycle basis to target small sets of 1 to 8 hosts of often recently discovered SNe Ia or to take advantage of an extended distance range from a new \textit{HST} instrument.  Near the end of the sample collection, the SH0ES Team targeted the completion of the sample to the above redshift limit, therefore it is difficult for the authors to resolve precisely which limit, distance or redshift, was the most relevant.} SH0ES Cepheid program proposals and target lists mix discussion of the redshift or distance limit. { Ultimately we conclude that we are unable to determine \textit{a priori} the selection unambiguously, so we model both and account for this ambiguity as a systematic uncertainty.}

\begin{figure}
    \centering
    \includegraphics{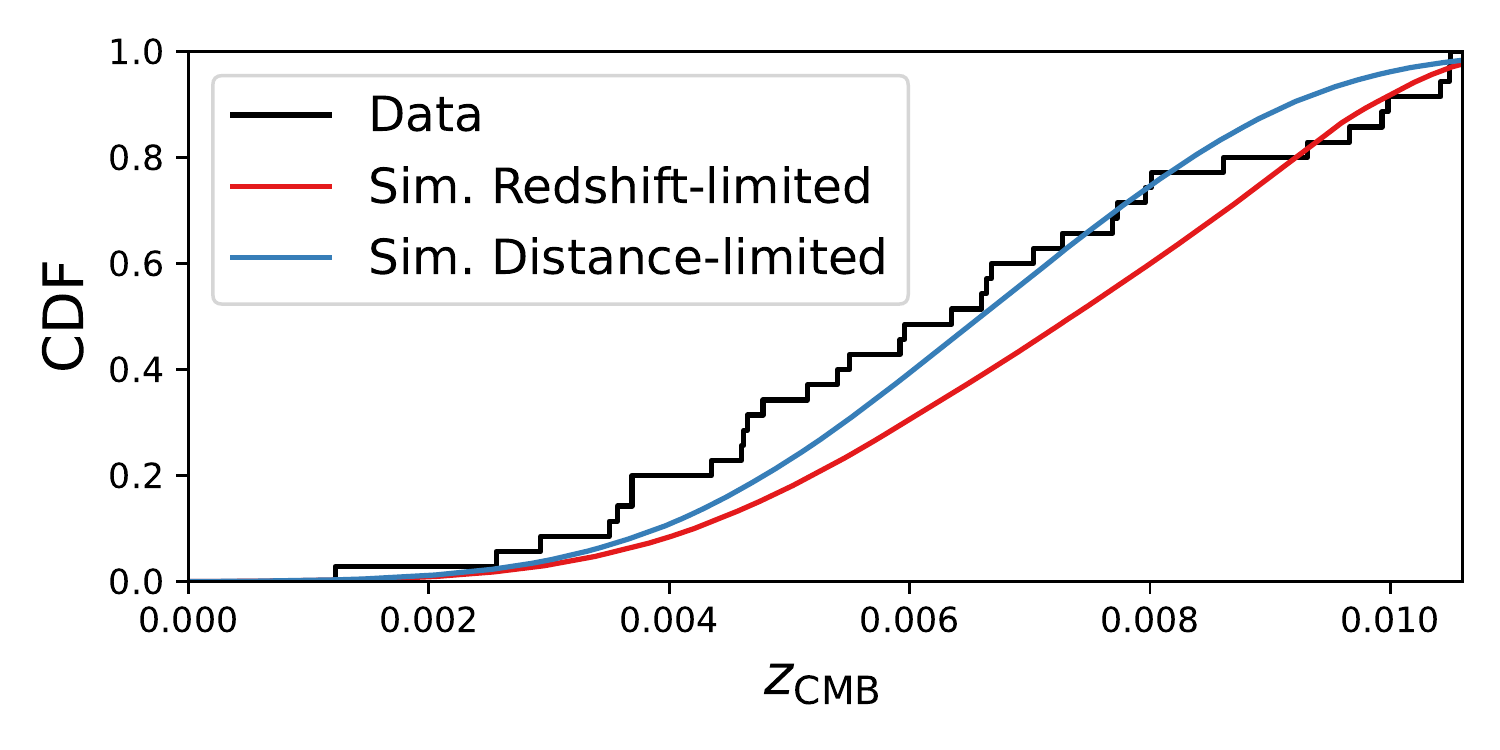}
    \caption{ The observed cumulative distribution function (CDF) of CMB frame group redshifts in the SH0ES sample is shown in black. The distribution is compared to simple Monte Carlo predictions for the observed distribution of redshifts if the SH0ES sample were to have been selected in a ``redshift-limited'' (red line) or a ``distance-limited'' scenario (blue line). {There is not evidence to reject the null hypothesis for either scenario.}} 
    \label{fig:zdistribution}
\end{figure}

\begin{figure}
    \centering
    \includegraphics{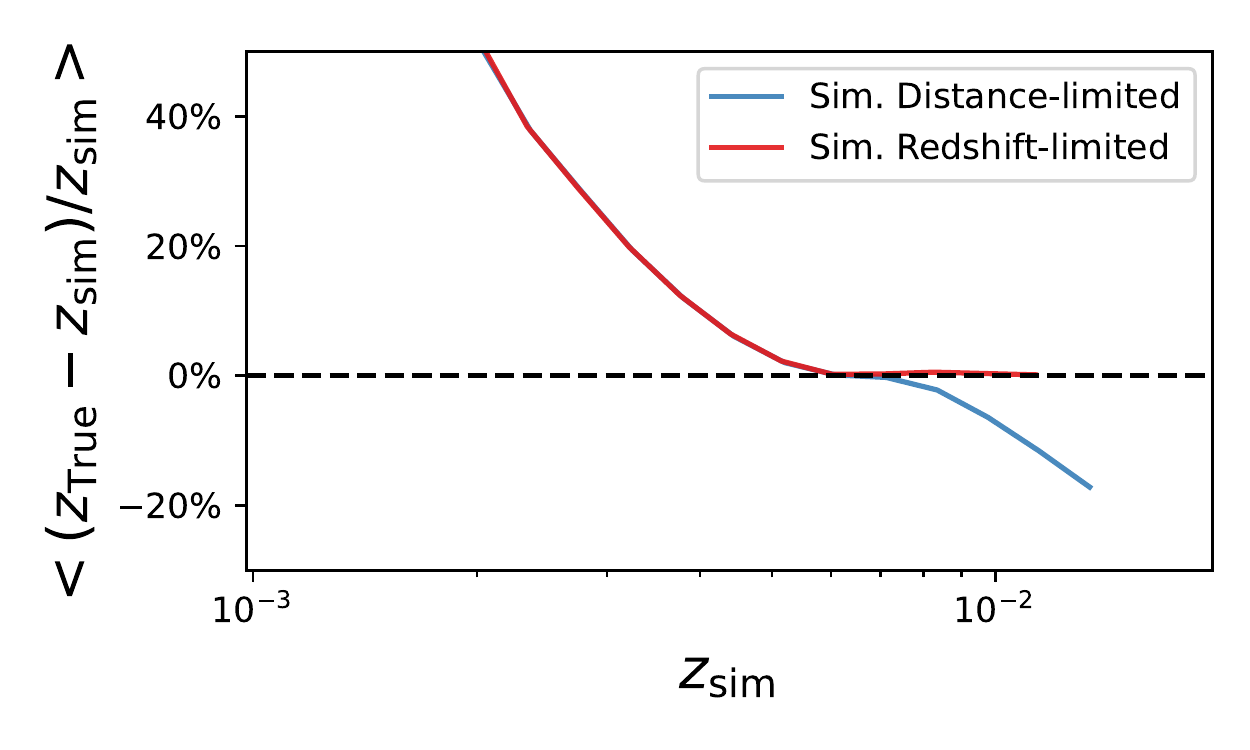}
    \caption{ Comparison of the fractional predicted bias in redshifts as a function of observed CMB redshift, derived from Monte Carlo simulations. Bias is shown if the SH0ES sample were to have been selected in a ``redshift-limited'' (red line) or a ``distance-limited'' scenario (blue line). In both scenarios at $z \lessapprox 0.008$, volumetric bias causes more SNe to have scattered down to their observed value than scattered up. At $z \gtrapprox 0.008 $, the predicted bias diverges, as galaxies at larger redshifts could only have made it into the distance-limited sample past a cut on SN distance by scattering up in redshift. }
    \label{fig:zbias}
\end{figure}

To better understand the impact of selection, we ran simple Monte Carlo simulations, in which we sampled galaxies from a 3 dimensional volume, using the density values found along SH0ES lines of sight in the \citet{Carrick2015} reconstruction to weight the distances. We simulated redshifts with a \SI{250}{
\kilo\meter\per\second} scatter, and SN distance moduli with 0.15 mag of scatter. We then compared the observed distribution of redshifts in the SH0ES sample to the simulated redshifts observed when using a strict cut on either redshift or SN distance in Figure \ref{fig:zdistribution} just above the most distant point in the SH0ES sample. We show the predicted bias as a function of simulated CMB redshift from these Monte Carlo simulations in Figure \ref{fig:zbias} for both scenarios. The bias is very similar for both at $z< 0.007$ but diverges at higher redshifts.

While the {redshift distribution of the SH0ES sample} can be seen to slightly disfavor selection by a redshift cut (a Kolmogorov-Smirnov test gives a p-value= 0.11), {neither scenario is inconsistent with the data.} As our data cannot rule out either scenario, we produce results assuming two limiting scenarios:

\begin{enumerate}
    \item \textbf{Assume redshift-limited selection}: We will assume that no external distance information of any kind was used in selection, besides the galaxy redshift. This assumption is more conservative as regards final uncertainties, as this reduces the information in our sample.
    \item \textbf{Assume distance-limited selection}: We assume a strict cut on SN\,Ia distance modulus was incorporated into the selection. As this assumption entails that the most distant redshifts in our sample are {are biased high compared to their true distances}, this assumption produces a lower value of the Hubble constant. This assumption is more conservative as regards the ongoing Hubble tension; alternative selection models consistent with the data would reduce the dependence of the selection on SN\,Ia information, reducing the impact.
\end{enumerate}

We account for other aspects of selection for SH0ES in addition to the distance criterion. As SN\,Ia progenitors are stellar remnants formed in galaxies, our analysis will use the galaxy density distributions found by \citet{Carrick2015} and \citet{LilowNusser2021ConstrainedFlow} to account for the inhomogeneous density of galaxies as a function of distance along each line of sight. Doing so additionally accounts for any bias due to the common practice (until the advent of real-time all-sky surveys) of targeting supernova searches to galaxies selected from catalogs rather than random pointings over the entire sky\footnote{Comparing, for example, the methodology of the Lick Observatory Supernova Search \citep{Leaman2011LOSSMethods} to that of ATLAS \citep{Tonry2018ATLAS}}.  We do not expect the star-forming galaxies which were then targeted by SH0ES to have galaxy bias parameters significantly different from unity \citep{Swanson2008SDSSstochasticity}, and thus expect the distribution of these galaxies to trace those of the reconstructions.   

Lastly, we have checked whether discovery bias could affect our results; i.e.,~whether that a SN\,Ia was reported before maximum light provides additional information about the distance of the host galaxy. This would be violated if the magnitude limit for discovery is fainter than the dimmest SN in our sample. For each SN we checked who was determined as discoverer by the IAUC\footnote{Center for Astronomical Telegrams, data available at \url{http://www.cbat.eps.harvard.edu/lists/Supernovae.html} or at \url{https://www.wis-tns.org/}}. For each person or collaboration who discovered a SN\,Ia in our sample, we determined the median magnitude of the supernovae they found at discovery. We find 32 out of 35 SNe were found by discoverers with a median discovery magnitude higher than the peak magnitude of the dimmest SN Ia in our sample
. We expect that discovery bias will not significantly affect our measurement, although it may more strongly affect selection at higher distances. {If discovery bias were to affect the sample, the effect on the measurement of $H_0$ would resemble the scenario we call ``distance-limited''.}

Given these considerations, the prior for the latent distances is proportional to the number density of galaxies per unit distance 
\begin{equation}
    P(\distance) \propto \distance^2  \times (1+\delta_\text{Rec.}( {\distance} \times \hat{n}_i)) \label{eq:redshiftdistanceprior}
\end{equation}
\noindent, {using the reconstructed density field  $\delta_\text{Rec.}(\Vec{r})$}. If no other distance information was incorporated into the SH0ES selection, this prior is sufficient. We use this definition of the distance prior in our analyses that ``Assume redshift-limited selection''.


If the SH0ES sample was selected using the information from the SNe\,Ia, then Equation \ref{eq:redshiftdistanceprior} should be modified to include the constraint that the observed SN\,Ia distance modulus was less than some critical value. To conduct this analysis without including the actual observed values of SN magnitudes, we adopt the simple model of SN magnitudes as normally distributed with dispersion \SI{0.15}{\mag} and fit the cutoff distance as a free parameter. The distance prior becomes

\begin{align}
    f(\distance,d_T) &= \distance^2  \times (1+\delta_\text{Rec.}( {\distance} \times \hat{n}_i)) \times \Phi( (\mu(\distance) - \mu(d_T) ) / .15 ) \\
    P(\boldsymbol{\distance}_i| \log d_T) &= \frac{f(\boldsymbol{\distance}_i, d_T) }{ \int_0^{80} f(x,d_T) dx} \;\;,\;\;\; \boldsymbol{\distance}_i \in (0, \SI{80}{\mega\parsec\per\h}]
    \label{eq:sndistanceprior}
\end{align}
\noindent where $\Phi(x)$ is the CDF function of the standard normal distribution, and $d_T$ is the assumed distance cutoff in units of \SI{}{\mega\parsec\per\h} for inclusion in the SH0ES sample. The upper bound on distances is done purely for performance reasons, and $\SI{80}{\mega\parsec\per\h}$ is well above the distances of any object in our sample.

In this scenario we also include a prior on the hyperparameter $\log d_T$
\begin{equation}
    \log d_T \sim \mathcal{N}(\log 30, 0.4).
\end{equation}

\noindent As we are unable to distinguish between these two models of selection \textit{a priori}, we present results using both priors. In Section \ref{sec:results} we discuss further which model is most appropriate for the data.



\subsection{Model Summary}
\label{subsec:modelsummary}

To conduct this measurement, we have defined a hierarchical model with 7 scalar parameters and a latent parameter for each Cepheid host galaxy, as well as one hyperparameter included in some analyses:
\begin{enumerate}
    \item The value of the Hubble constant  $
    \log H_0 $, in units of \SI{}{\hubbleunit}. This is our only parameter of interest, and the rest of our parameters may be regarded as nuisance parameters included to quantify related systematic uncertainties.
    \item $\beta$, the scale parameter that relates galaxy overdensity to peculiar velocities.
    \item $V_\text{Ext}$, the magnitude of the bulk flow external to the volume of the reconstruction in units of \SI{}{\kilo\meter\per\second} and $(l,b)$ the galactic longitude and latitude of its direction. 
    \item $\boldsymbol{\distance }$, the latent comoving distance of each Cepheid host galaxy to the Milky Way in units of Mpc $h^{-1}$.
    \item Two scale parameters $\sigma^\mathrm{(PV)}_\mathrm{Unmodelled}, \sigma^\mathrm{(PV)}_\mathrm{Structure} $, which control the amplitude of the peculiar velocity uncertainties on scales below/above  $\sim \SI{7}{\mega\parsec\per\h}$ respectively.
    \item When assuming distance-limited, a hyper-parameter $\log d_T$ which sets the assumed cutoff of the distance-limited sample (i.e. the maximum SN\,Ia magnitude allowed to pass cuts). 
\end{enumerate}

Using this methodology allows us to naturally budget for the majority of our systematic effects, account for the selection of our sample, and account for the off-diagonal effects of correlated large scale structure.

When applied to our data, the data strongly dominates the posterior distribution over $\log H_0$, while the prior largely determines the posterior distribution over the parameters of the velocity reconstructions. This is as expected, as with only 35 data points we do not expect to constrain the velocity field in the local volume as a whole, and rely on the values of these parameters measured in the original work. When fit $\sigma^\mathrm{(PV)}_\mathrm{Unmodelled}$ is mostly well constrained, while $\sigma^\mathrm{(PV)}_\mathrm{Structure}$ is well constrained if ``distance-limited'' selection is assumed. $\log d_T$ is strongly constrained.

\section{Analysis and Implementation}
\label{sec:implementation}
\subsection{Computation}

We use the code Stan \citep{Carpenter2017Stan} to sample from the posterior distribution of the parameters as defined by the prior and likelihoods given above. Accounting for the full dependence of the velocity covariance on the latent distances within {Stan}'s HMCMC framework is computationally intractable, requiring evaluation by chain rule of the derivatives of the likelihood with respect to $\sim 1200$ matrix elements for $\sim 35$ parameters. As the effect is small, we use the MCMC to sample from the posterior while keeping the velocity covariance fixed, evaluated at fiducial values for the distances, then use importance sampling to reweight each sample appropriately. Given the posterior samples of the parameters $\theta^{(i)}$, the relative weight of the $i$th sample is given by the ratio of the likelihoods of the observed redshifts with the covariance varying/fixed
\begin{equation}
   w^{(i)}=\frac{ p(\boldsymbol{z}_\text{CMB} |\theta^{(i)}, \boldsymbol{\Sigma}(\boldsymbol{\distance}^{(i)}))}{p(\boldsymbol{z_\text{CMB}}|\theta^{(i)}, \boldsymbol{\Sigma}_\text{Fid})}.
\end{equation}
The overall effect of this correction is small; typically $\sim 0.1\sigma$ in $H_0$.

\subsection{Effects of each component of modeling}
\label{subsec:modeldiffs}

To show why the complexity of the model we have constructed is required, we break down how each component of the model affects the central value and error estimate of our result in Table \ref{tab:modeldiffs}. We use the following simplifications of our model:
\begin{enumerate}
    \item \textbf{Simple Estimator}: Measurement of $H_0$ from the weighted mean of $5 \log H_0= 5 \log_{10}(c z_\mathrm{CMB}) - \mu_\mathrm{Obs.}$, with \SI{250}{\kilo\meter\per\second} of peculiar velocity noise and diagonal errors from the Cepheid uncertainties $\boldsymbol{\Sigma}^{(\mathrm{SH0ES})}_\mu$. {No PV corrections are applied to the redshifts.} This gives a central value of $\SI{ \hubblevaluesimpleestimatorcentral}{
    \hubbleunit}$.
    \item \textbf{Include Cepheid Systematics}: We use the hierarchical model based on Equations \ref{eq:mulikelihood} and \ref{eq:zlikelihood} without PV corrections, and include the full systematic covariance matrix for the Cepheid measurements (see Appendix \ref{app:cephsys}).
    \item \textbf{Include  PV Corrections}: We include the PV corrections discussed in Section \ref{subsec:pvpred}, based on the \citetalias{Carrick2015} reconstructions.
    \item \textbf{Include PV Covariance Estimate}: We include the full estimate of PV covariance for corrected redshifts discussed in Section \ref{subsec:pvcovariance}.
     \item \textbf{Include Selection Effects (distance-limited) (Baseline model)}: We account for the model of distance-limited selection discussed in Section \ref{subsec:selectionpriors}, using the prior in Eq. \ref{eq:sndistanceprior}. This is the same result presented in the 7th row of Table \ref{tab:results}.
    \item \textbf{Simple Estimator, $cz>\SI{2000}{\kilo\meter\per\second}$}: As the Simple Estimator, but removing the data at $cz<\SI{2000}{\kilo\meter\per\second}$. This reduces the impact of several systematic uncertainties that are otherwise addressed by our modeling. This reduces the sample size from 35 objects to 15. This estimator gives a central value of $\SI{\hubblevaluesimpleestimatorcutcentral }{
    \hubbleunit}$.
\end{enumerate}
We see that ignoring any component of the model will substantially bias the inferred value of the Hubble constant, and/or lead to underestimates in the inferred error budget. Note that the first and second rows of Table \ref{tab:modeldiffs} underestimate cosmic variance and do not account for selection effects, and should not be used outside this context; results which do not make use of any local PV models are shown in the first two rows of Table \ref{tab:results}.

\begin{table}[]
    \centering
    \textbf{Table 1: Two-Rung Simple Estimator and Cumulative Analysis Changes}\medskip
    \begin{tabular}{lrrl}
\hline
                                                                                   &   $ H_0 - \SI{ 72.8}{\kilo\meter\per\second\per\mega\parsec}$ &   $\sigma H_0$ & $\Delta H_0$ from last   \\
\hline
 Simple Estimator                                                                  &                                                          -2.5 &            1.7 & -                        \\
 Include Cepheid systematics                                                       &                                                          -3.2 &            1.9 & -0.7                     \\
 Include PV Corrections                                                            &                                                          -0.6 &            2   & 2.6                      \\
 Include PV Covariance Estimate                                                    &                                                           1.5 &            2.7 & 2.1                      \\
 Include Selection Effects (distance-limited) (see Row 7, Table \ref{tab:results}) &                                                           0   &            2.6 & -1.5                     \\
 \hline Simple Estimator, $cz> \SI{2000}{\kilo\meter\per\second}$                  &                                                          -0.6 &            2.1 & -                        \\
\hline
\end{tabular}
    \caption{{Effect of components of the model, as discussed in Section \ref{subsec:modeldiffs}, on the $H_0$ measurement. Except for the last row, we cumulatively add each element of the model to show the impact of each element on the error estimate and central value of $H_0$. The full model, as shown in the second to last row, is used to produce the full results in Row 7 of Table \ref{tab:results}; it uses the \citetalias{Carrick2015} reconstruction with assumed distance-limited selection and fixed PV amplitude parameters.  The last row, given for reference, shows the Simple Estimator when only galaxies with $cz > 2000$ km s$^{-1}$ are used. 
}}
    \label{tab:modeldiffs}
\end{table}

\subsection{Validation} \label{subsec:validationreal}

To validate our approach, we wish to ensure that the model is capturing the features found in the real data, as well as compare the use of the \citetalias{Carrick2015} and \citetalias{LilowNusser2021ConstrainedFlow} reconstructions to a more conservative analysis which does not make use of PV information from sources external to SH0ES. To do so we use leave-one-out (``loo'') cross-validation based on \citet{vehtari2017practical} and \citet{burkner2021efficient} to quantify the performance of different models. By refitting the model while withholding one data point from the model, we determine which model most accurately predicts the held-out data. Our metric is the expected log predictive density 
\begin{align}
    \widehat{\mathrm{elpd}}_\mathrm{loo}^\mathcal{M} & = \sum_{i=1}^N \log p(\boldsymbol{\mu}_i | \boldsymbol{\mu}_{-i},\mathcal{M}) \\
    p(\boldsymbol{\mu}_i | \boldsymbol{\mu}_{-i},\mathcal{M}) &= \int p(\boldsymbol{\mu}_i| \boldsymbol{\mu}_{-i},\theta,\mathcal{M}) p(\theta| \boldsymbol{\mu}_{-i},\mathcal{M}) d \theta
\end{align}
where $ \boldsymbol{\mu}_{-i}$ is the vector of distance moduli with the $i$th entry removed, and $\mathcal{M}$ denotes the choice of peculiar velocity reconstructions  and other modelling choices. 
. Higher $\widehat{\mathrm{elpd}}_\mathrm{loo}$ indicates that the model does a better job at predicting the left out data.  As a log-likelihood based metric, elpd can be interpreted similarly to similar metrics such as the Bayesian evidence. Absolute values of this quantity are not informative in themselves, but the differences between models allow discrimination between them. We compare the elpd to a reference, which we choose as the model which uses no SH0ES-external information, assumes distance-limited selection, and fits the amplitude of PV covariance. As the estimated elpd is a summation of $N$ independent terms, we estimate the significance of these differences by bootstrapping the estimate to evaluate a p-value which we write as  
\begin{equation}
    p(\mathrm{elpd}<\mathrm{Ref.})= p({\mathrm{elpd}}^\mathcal{M} < {\mathrm{elpd}}^ \mathrm{Ref.}) 
\end{equation}

\noindent which quantifies the likelihood that a given analysis would produce an elpd better than that of the reference model by chance. Small p-values indicate that the reconstructions used are informative about the underlying peculiar velocities.  

\begin{figure}
    \centering
    \includegraphics{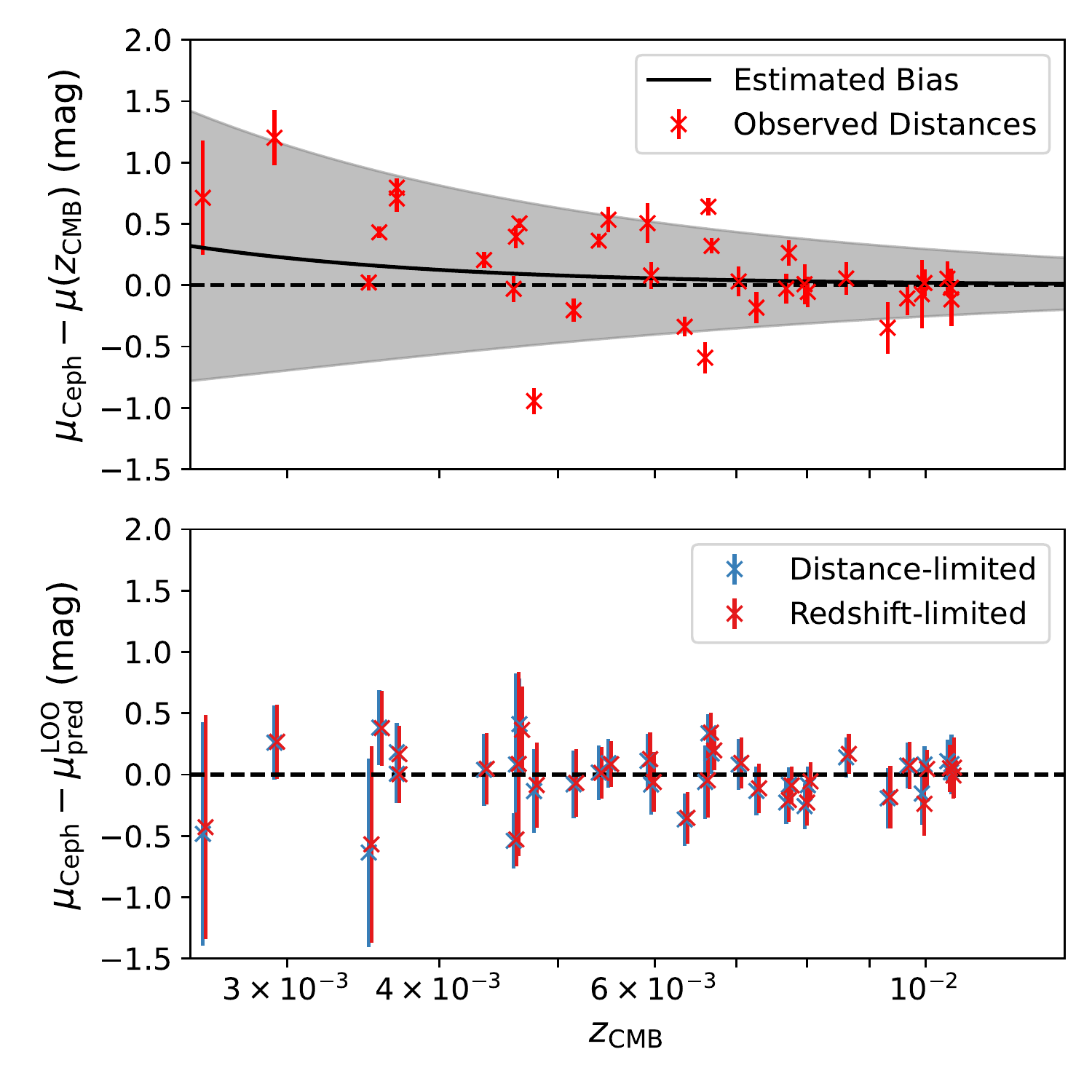}
    \caption{Upper Panel: Observed Cepheid distance moduli, with a smooth Hubble flow subtracted based on naive CMB-frame redshifts. We show the Malmquist bias predicted from \SI{250}{\kilo\meter\per\second} of diagonal scatter in black. Gray shading shows the amplitude of correlated peculiar velocities expected from $\Lambda$CDM.  Structure can clearly be seen by eye in the residuals. {We show the expected bias from a redshift-limited sample in black. } Lower Panel: Residuals of leave-one-out tests, where observed distance moduli are compared to the distances predicted from redshifts, the \citetalias{Carrick2015} reconstruction, and the distances of the rest of the sample. Amplitude of PV covariance is fitted, and we show results assuming both distance-limited and redshift-limited selection. Error bars show the dispersion in the predicted distances for each point. There is no significant linear trend, and no significant autocorrelation ($r<0.2$) at any lag.}
    \label{fig:residualcomparison}
\end{figure}

\begin{figure}
    \centering
    \includegraphics{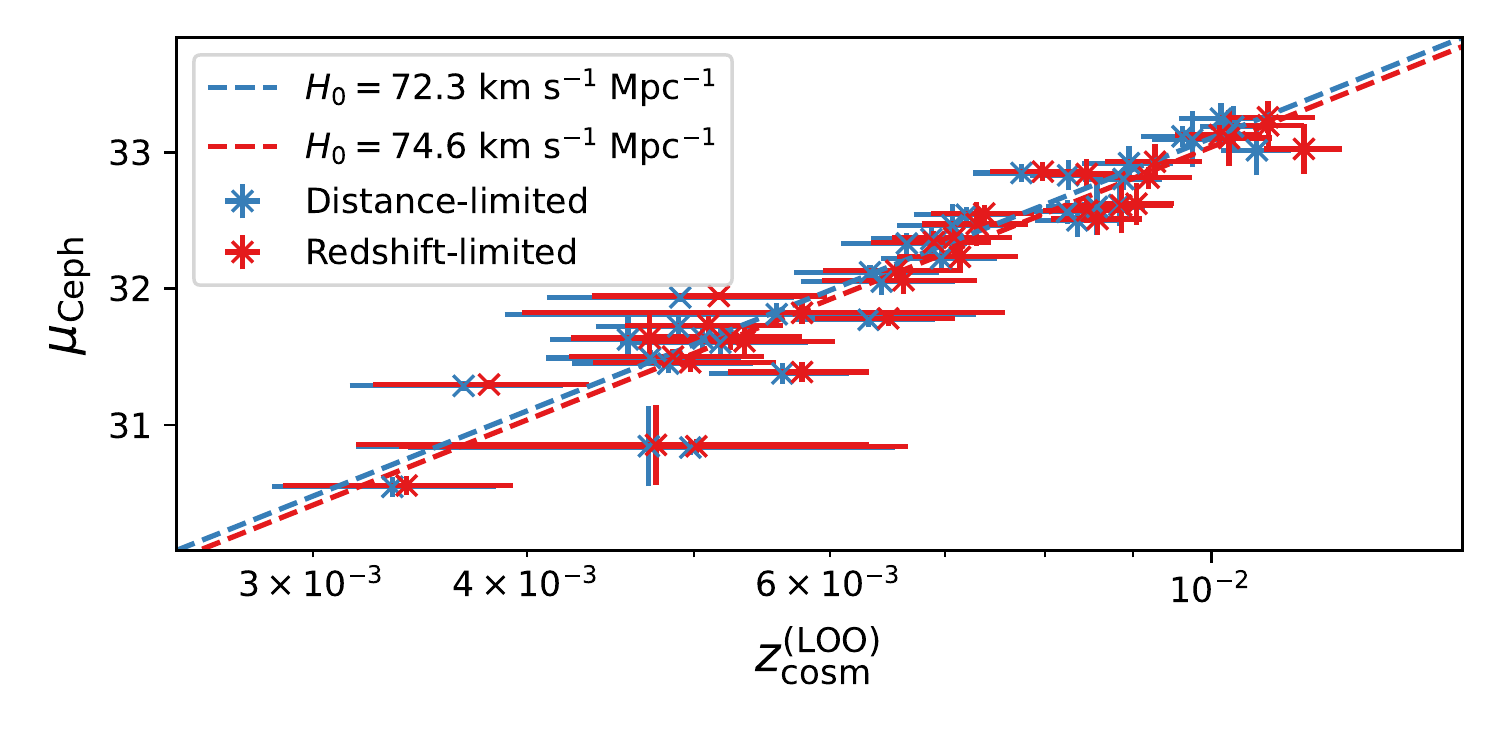}
    \caption{ Hubble diagram derived from leave-one-out tests. We show observed Cepheid distances on the y-axis, and the predicted cosmological redshift inferred from the redshift of the galaxy corrected for bias and peculiar velocities on the x-axis. These predicted redshifts are determined while the Cepheid distance for each point is held out. Modeling assumes \citetalias{Carrick2015} reconstruction, and fits the amplitude of peculiar velocities. We show results assuming both distance-limited and redshift-limited selection. The point with large $z$-uncertainties is NGC 4424, whose observed redshift places it in the middle of Virgo cluster, and whose cosmological redshift is highly uncertain as a result.}
    \label{fig:hubblediag}
\end{figure}

We can also examine the residuals of the leave-one-out tests by directly comparing the observed value of the distance moduli $\boldsymbol{\mu}_\mathrm{Obs.}$ to the predicted values when that data point is left out of the model fitting. As the predictions incorporate the information of the rest of the sample, these residuals should be uncorrelated (and approximately normal in distribution). We show a comparison of the residuals relative to the predictions in Figure \ref{fig:residualcomparison}, and a comparison of predicted redshifts to observed distance moduli in Figure \ref{fig:hubblediag}. The positive tail in the uncorrected residuals seen in the upper panel of Figure \ref{fig:residualcomparison} at low distances is a distinctive artifact of the volumetric or Malmquist bias due to scatter in redshifts. The same artifact at these redshifts can be seen in the SN\,Ia Hubble residuals of the Pantheon+ analysis \citep{Brout2022PantheonPlus}. These data were not used in the Pantheon+ cosmology fit, which cut the sample at $z=0.01$, nor in the SH0ES analysis \citetalias{Riess2021ComprehensiveSH0ES} which cut the sample at $z=0.023$. The bias is not expected to affect either result, as it scales as $z^{-2}$ and is negligible in both samples.


\section{Results}

\label{sec:results}

We first measure the Hubble constant without the use of peculiar velocity reconstructions.  This gives us a measurement independent of PV reconstruction procedures, and allows us to quantify the improvement from the use of this data. {We use the full $\Lambda$CDM power spectrum as described in Equation \ref{eq:nocorrpowerspec}}, set $v_\mathrm{pred}(\boldsymbol{\distance}_i)$ and $\delta(\boldsymbol{\distance}_i)$ to 0 in Equations \ref{eq:zlikelihood} and \ref{eq:sndistanceprior}, and remove the reconstruction parameters $\vext,l,b,\beta$ from the model. Without the use of other peculiar velocity information , we measure $H_0$ to be $ \hubblevaluefixpvcovsncutcosmowitherr\ \SI{}{\hubbleunit}$, a precision of $\sim 6.3\%$. {Fitting the peculiar velocity amplitudes leads to a smaller estimate of the small-scale or nonlinear power, yielding $H_0=\hubblevaluefitpvcovsncutcosmowitherr\ \SI{}{\hubbleunit}$. This choice is significantly favored by the data. We take the latter model as the reference for a two-rung measurement without the use of PV information.} We show the values for $H_0$ and the other parameters in Table \ref{tab:results}. Ignoring the galaxy catalog information in the case of an assumed redshift cut produces pathological behavior in the posterior prior due to the lack of constraining power in redshifts, and we do not produce results for this scenario. 

\begin{table}[]
    \centering
    \textbf{Final Results: Variations on Selection and Peculiar Velocity Assumptions}\medskip
    \begin{tabular}{llllllll}
\hline
 Pec. Vel. Map                                      & Assumed Selection   & $H_0$                                 & $d_T$                    & $\sigma^\mathrm{(PV)}_\mathrm{Unmodelled}$   & $\sigma^\mathrm{(PV)}_\mathrm{Structure}$   & $\widehat{\mathrm{elpd}}_\mathrm{loo}$   & $p(\mathrm{elpd<Ref.})$   \\
\hline
                                                    &                     & (km s$^{-1}$ Mpc$^{-1}$)              & (Mpc $h^{-1}$)           & (km s$^{-1}$)                                & (km s$^{-1}$)                               & log-like                                 &                           \\
 \hline None                                        & Distance-limited    & $71.0_{ -4.6 }^{ +4.7 }$              & $28_{ -2 }^{ +2 }$       & Fixed to 264                                 & Fixed to 266                                & -6.57                                    & 0.952                     \\
 None                                               & Distance-limited    & $71.5_{ -4.5 }^{ +4.6 }$              & $29_{ -2 }^{ +2 }$       & $162_{ -35 }^{ +41 }$                        & $290_{ -69 }^{ +88 }$                       & -3.83                                    & Ref.                      \\
 \hline \citetalias{LilowNusser2021ConstrainedFlow} & Distance-limited    & $72.3_{ -2.7 }^{ +2.6 }$              & $29.6_{ -1.4 }^{ +1.5 }$ & Fixed to 264                                 & Fixed to 66                                 & -4.16                                    & 0.579                     \\
 \citetalias{LilowNusser2021ConstrainedFlow}        & Redshift-limited    & $76.5_{ -2.5 }^{ +2.5 }$              & -                        & Fixed to 264                                 & Fixed to 66                                 & -2.70                                    & 0.337                     \\
 \citetalias{LilowNusser2021ConstrainedFlow}        & Distance-limited    & $\boldsymbol{71.8_{ -1.9 }^{ +2.1 }}$ & $29.9_{ -1.2 }^{ +1.3 }$ & $136_{ -24 }^{ +29 }$                        & $65_{ -19 }^{ +29 }$                        & 1.24                                     & $\boldsymbol{0.015}$      \\
 \citetalias{LilowNusser2021ConstrainedFlow}        & Redshift-limited    & $\boldsymbol{74.2_{ -2.3 }^{ +2.7 }}$ & -                        & $137_{ -25 }^{ +31 }$                        & $77_{ -26 }^{ +45 }$                        & 1.67                                     & $\boldsymbol{0.009}$      \\
 \hline \citetalias{Carrick2015}                    & Distance-limited    & $72.7_{ -2.6 }^{ +2.6 }$              & $29.7_{ -1.3 }^{ +1.5 }$ & Fixed to 264                                 & Fixed to 66                                 & -3.21                                    & 0.415                     \\
 \citetalias{Carrick2015}                           & Redshift-limited    & $77.0_{ -2.6 }^{ +2.5 }$              & -                        & Fixed to 264                                 & Fixed to 66                                 & -1.84                                    & 0.203                     \\
 \citetalias{Carrick2015}                           & Distance-limited    & $\boldsymbol{72.3_{ -1.9 }^{ +2.0 }}$ & $30.1_{ -1.2 }^{ +1.2 }$ & $132_{ -23 }^{ +27 }$                        & $63_{ -18 }^{ +26 }$                        & 2.26                                     & $\boldsymbol{0.008}$      \\
 \citetalias{Carrick2015}                           & Redshift-limited    & $\boldsymbol{74.4_{ -2.2 }^{ +2.5 }}$ & -                        & $132_{ -24 }^{ +28 }$                        & $72_{ -24 }^{ +39 }$                        & 2.93                                     & $\boldsymbol{0.002}$      \\
\hline
\end{tabular}\medskip
\centering
\textbf{Fiducial Result:    }$73.1_{ -2.3 }^{ +2.6 }\ \SI{}{\kilo\meter\per\second\per\mega\parsec}$ \medskip

    \caption{Posterior medians, with upper and lower uncertainties from the 16th and 84th percentiles, of our parameter of interest $H_0$ and nuisance parameters under permutations of three analysis choices: choice of PV reconstruction, assumed selection of the SH0ES sample, and whether to fix or fit the amplitude of peculiar velocity covariance. {The four bolded values of $H_0$ are stacked to produce the fiducial result.} Final columns show the expected log predictive density (elpd) and associated p-value; {large elpd's and small p-values indicate better information on peculiar velocities.  See Section \ref{subsec:validationreal} for details.}
}
    \label{tab:results}
\end{table}

\begin{figure}
    \centering
    \includegraphics{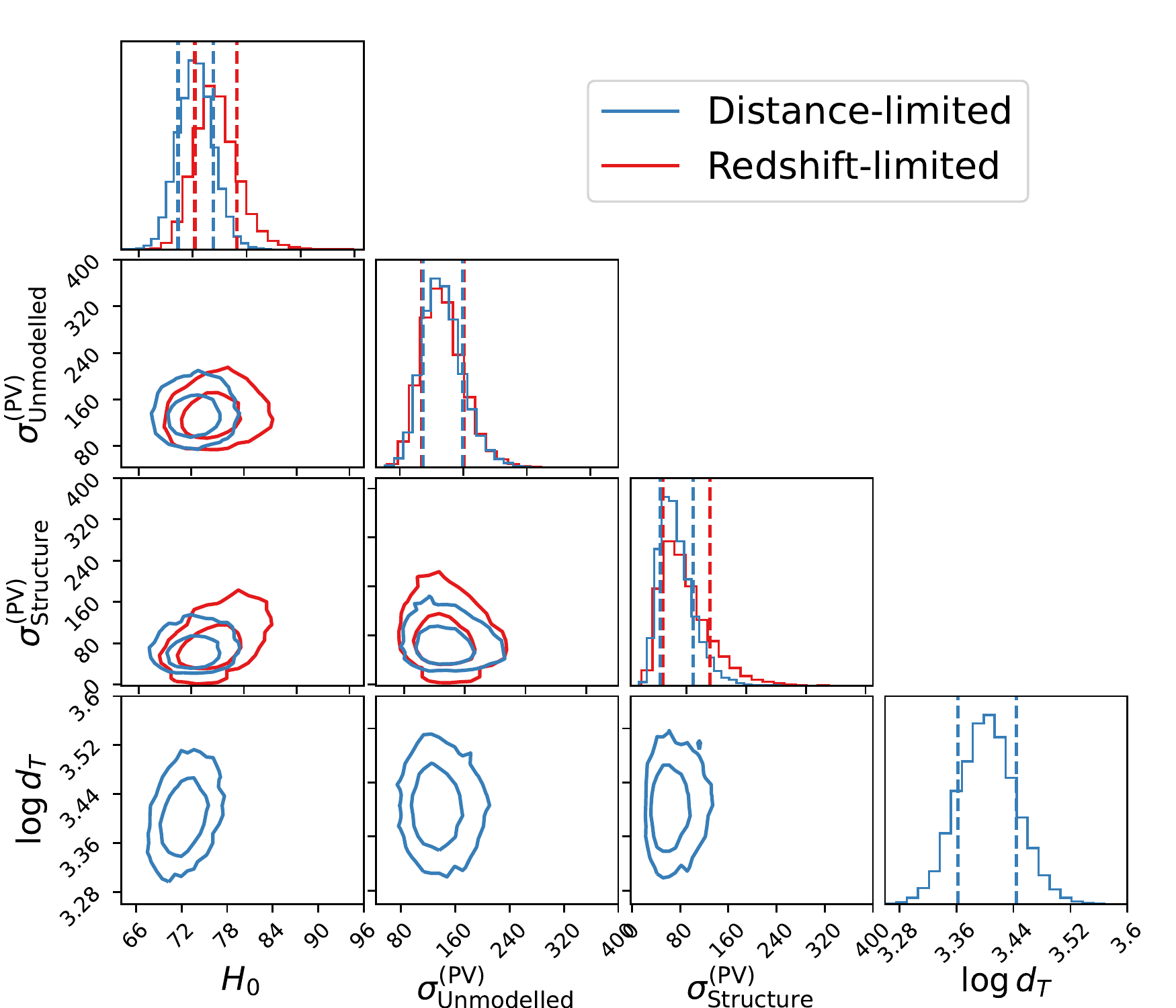}
    \caption{Corner plot showing the posterior distribution of the scalar parameters given the observed redshifts, distance moduli, and the \citetalias{Carrick2015} reconstruction, fitting the PV amplitude. Contours of parameters assuming both distance-limited and redshift-limited selection are shown in blue and red respectively. We show 68\% and 95\% contours for the 2d histograms, with 16\% and 84\% quantiles for the 1d histograms. $H_0$ shows degeneracy with $\log d_T$ in distance-limited case, and $\sigma_\mathrm{Structure}^{\mathrm{(PV)} }$ in redshift-limited case, showing the extent to which selection uncertainties limit our measurement.}
    \label{fig:carrick_cornerplot}
\end{figure}

We then fit the data using each of the two reconstructions, two choices of prior, and the choice to either fit the amplitude of peculiar velocity covariances or fix them to fiducial values. We report the appropriate parameter constraints in Table \ref{tab:results}.  Including information from the reconstructions, the fractional error in $H_0$ decreases from $\sim 6\%$ to $\sim 3\%$, with the assumption of SN Ia selection decreasing uncertainties further. We show the corner plot of the constraints on the scalar parameters using the \citetalias{Carrick2015} reconstruction with fitted PV covariance in Figure \ref{fig:carrick_cornerplot} under both selection scenarios. 

As expected, use of either reconstruction is significantly favored by the elpd metrics. The reconstruction of \citetalias{Carrick2015} is slightly favored by the data compared to \citetalias{LilowNusser2021ConstrainedFlow}, but the difference is not significant. The choice to fit the peculiar velocity amplitudes is also strongly favored by the data. The data slightly favors the redshift-limited scenario for selection, but the difference is not significant ($p>0.2$ for both \citetalias{LilowNusser2021ConstrainedFlow} and \citetalias{Carrick2015} when peculiar velocity amplitudes are fitted). { We conclude that we are incapable of differentiating the models of selection from the data alone. As the redshift-limited scenario implies a more restricted experimental process, we consider both equally likely given the data.} 

Inclusion of the reconstructions shifts the central value by $\sim  \SI{1.5}{\hubbleunit}$, much less than the $\sim \SI{3}{\hubbleunit}$ reduction in the uncertainty, and thus within expectation. We note that the two reconstructions produce very similar results regardless of the other analysis choices made, consistent with our error budget. The different approaches to selection shift results by more, increasing the value of $H_0$ by $\sim 2 \text{--} \SI{5}{\hubbleunit}$, depending on whether or not the peculiar velocity covariance is also fit. 

The fitted values of the amplitude of PV covariance are consistent with our expectation. As we have used group-corrected redshifts, we expect that much of the diagonal scatter has been removed, lowering the value of $\sigma^\mathrm{(PV)}_\mathrm{Unmodelled}$; our fitted values are consistent with the estimated \SI{150}{\kilo\meter\per\second} for galaxy halos found by comparison to N-body simulations in \citet{Carrick2015}, and assumed in \citet{Betoule2014}, \citet{Boruah2020CosmicConstraints}, and others. However, \citet{Brout2022PantheonPlus} finds a peculiar velocity scatter of \SI{240}{\kilo\meter\per\second} in the Pantheon+ supernova sample. This number was found in a mixed sample of galaxies with and without group-assignments, at larger distances where group assignments are more uncertain. {Additionally PV corrections for the baseline Pantheon+ analysis were not integrated over the line of sight, which tends to increase statistical dispersion in distances \citep{Peterson2021PVCorrections}}. The values of $\sigma^\mathrm{(PV)}_\mathrm{Structure}$ are in all cases consistent with their fiducial values, showing that our estimates of the magnitude of errors in the reconstructions are reasonable and that the scale of observed local structures is consistent with that expected from $\Lambda$CDM.

\begin{figure}
    \centering
    \includegraphics{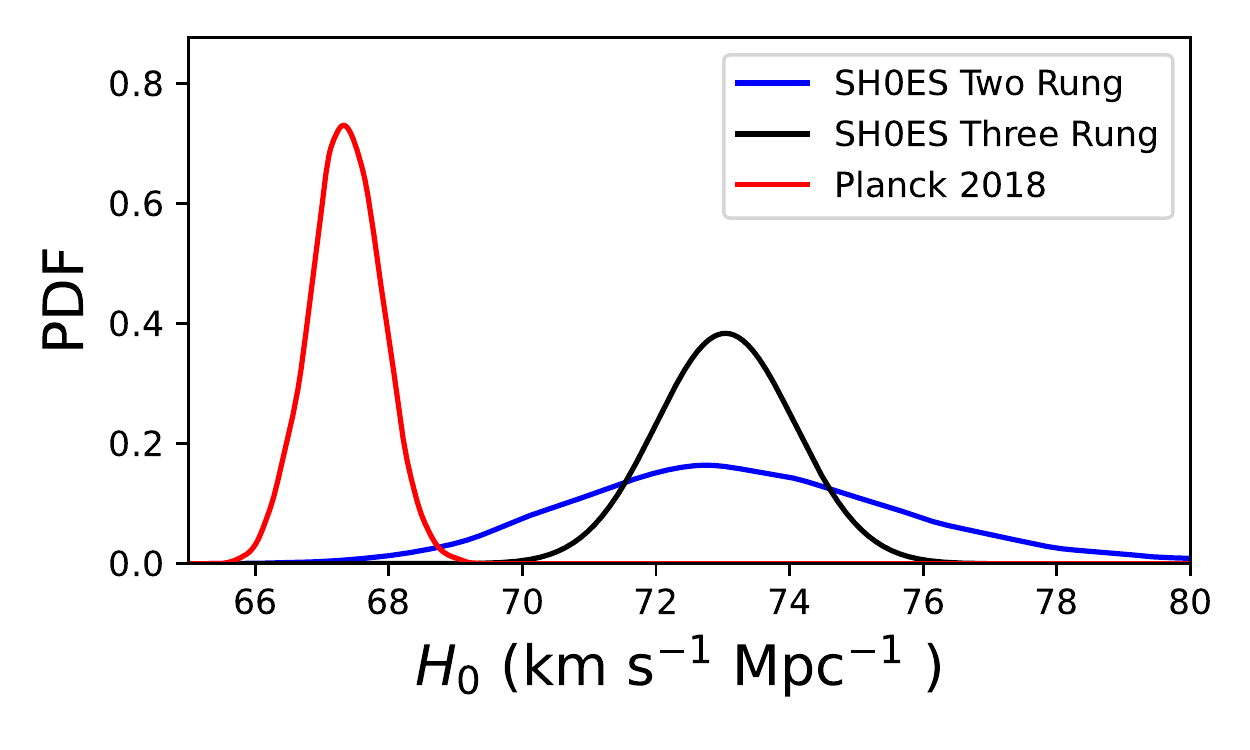}
    \caption{Comparison of our measurement of $H_0$ from a two rung measurement using Cepheids and geometry to the prediction of \citet{PlanckCollaboration2018} and the measurement from the SH0ES three rung distance ladder \citep{Riess2021ComprehensiveSH0ES}. We find  that our result is consistent with the full SH0ES value, and at $2.6\sigma$ tension with inferred values of $H_0$ from Planck.}
    \label{fig:final_results}
\end{figure}

As the elpd strongly favors the use of peculiar velocity reconstructions and strongly favors fitting of the peculiar velocity amplitudes, we stack the posterior estimates of $H_0$ from these four models, assigning equal weights to each. The resulting median value for $H_0$ is  $H_0=\hubblevaluefiducialwitherr\ \SI{}{\hubbleunit} $ (with 16\% and 84\% percentile uncertainties).  We take this as our fiducial result. We evaluate a $p$-value for the difference between our value and the Planck measurement of $H_0=67.4 \pm 0.5\ \SI{}{\hubbleunit} $ {\citep{PlanckCollaboration2018}}. Comparing the MCMC chains, we find $p=0.004$, corresponding to a significance of $2.6\sigma$. We compare our result to \citetalias{Riess2021ComprehensiveSH0ES} and \citet{PlanckCollaboration2018} in Figure~\ref{fig:final_results}.
 
\subsection{Comparison to SNe\,Ia Hubble Diagram}

\begin{figure}
    \centering
    \includegraphics{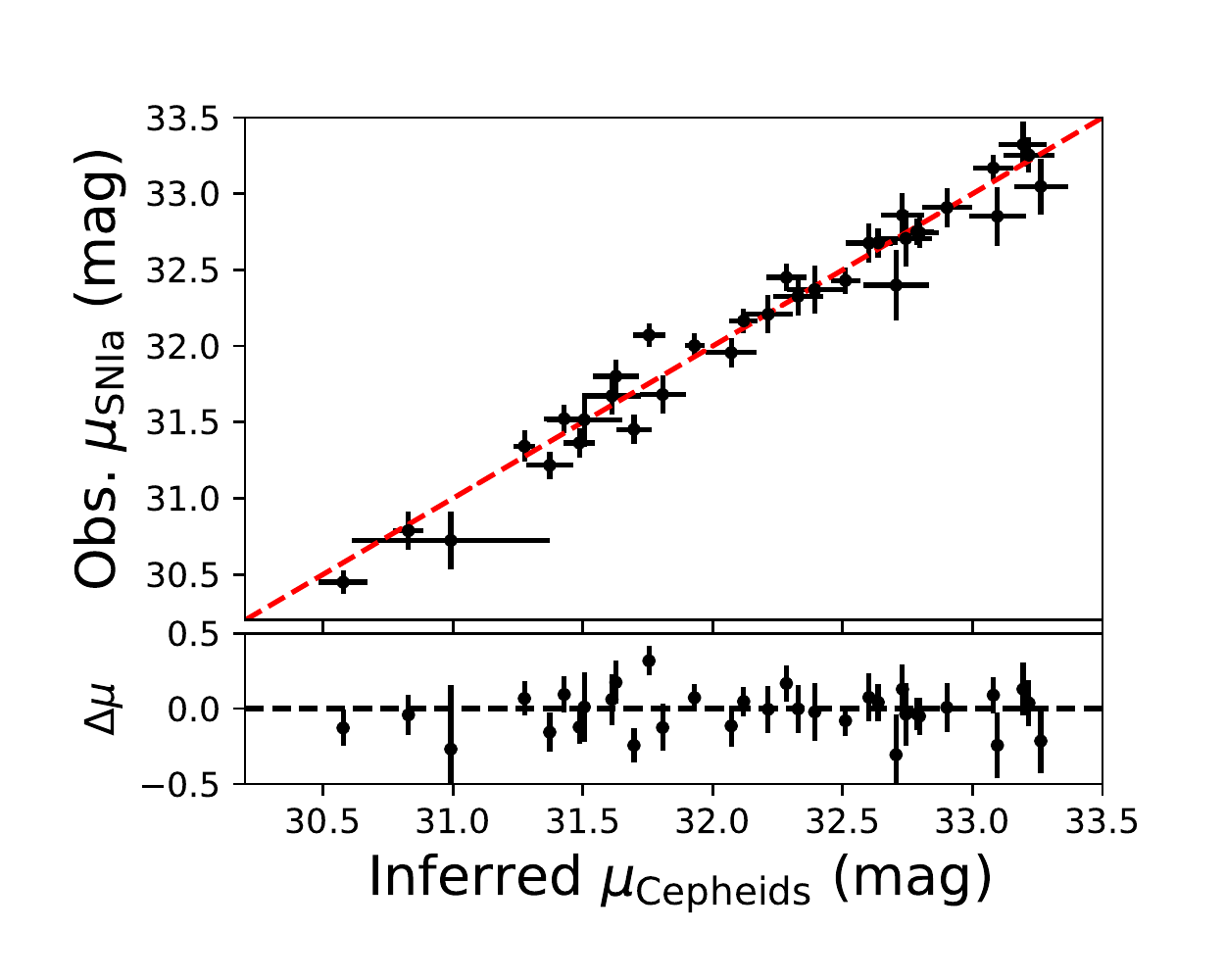}
    \caption{Comparison of observed SN\,Ia standardized magnitudes to the inferred distances measured from the combination of SH0ES Cepheid, redshifts, and the \citetalias{Carrick2015} reconstruction, fitting the peculiar velocity covariance, and assuming distance-limited selection. {Where a galaxy is host to multiple SNe\,Ia, we take a mean of their magnitudes, weighted by their covariance.}}
    \label{fig:snia_cross_comparison}
\end{figure}

As an additional crosscheck, we compare the distances we measure from our models to the Type Ia supernova Hubble diagram. We use the standardized magnitudes of 73 light-curves of cosmological SNe from the Pantheon+ sample \citep{Scolnic2021PantheonPlusLightcurves,Brout2022PantheonPlus} with common host galaxies to our SNe. To evaluate a goodness-of-fit metric, we compute a $\chi^2$ by comparing predicted values of $\mu$ from the model to SN magnitudes, fitting the absolute magnitude of the Type Ia supernovae, and including the covariance of the predicted $\mu$'s and the covariance of the supernovae provided by \citet{Brout2022PantheonPlus}. $\chi^2$ values do not vary significantly from model to model, between 49.4-51.0, with 72 degrees of freedom. That the values do not change significantly is not surprising, as the Cepheid distances have the least statistical error of any indicator used in this analysis, and the predicted distances are largely driven by the Cepheid measurement. We show a comparison of these two rungs of the distance ladder in Figure \ref{fig:snia_cross_comparison}. Our results are consistent with the results of \citetalias{Riess2021ComprehensiveSH0ES}.

\section{Conclusion}
\label{sec:conclusions}
We find that the Hubble constant can be constrained at extremely low redshifts by taking advantage of measurements of the local density and velocity fields from external sources. {Our code and data will be made public as of the acceptance of this paper.} Values of the Hubble constant measured when incorporating this information, depending on further modelling assumptions, range from \SI{\hubblevaluefitpvcovsncutLilowcentral}{\hubbleunit} to \SI{\hubblevaluefixpvcovzcutCarrickcentral}{\hubbleunit}. Our data supports a credible interval of $ H_0=\hubblevaluefiducialwitherr \SI{}{\hubbleunit}$, at $2.6\sigma$ tension with the measurement of Planck. As a result, we conclude the cause of the Hubble tension is unlikely to be an unknown systematic of Type Ia supernovae.

Our analysis is sensitive to the modelling of the Cepheid variable stars, including reddening, crowding, and metallicity corrections. However any systematic errors or uncertainties in this stage of the analysis will be mitigated by our inclusion of a covariance matrix derived from the analysis variants of \citetalias{Riess2021ComprehensiveSH0ES}.

We find that the most significant uncertainty of our analysis is the selection function of the SH0ES sample. Seeking a further reduction of the Hubble constant would require informative distance priors with increased probability mass at low distances. However as our data show good consistency with the observed distributions of distances, increasing the probability of low distances will require reductions in the information of the prior.  Reducing the size of the peculiar velocity beyond $\SI{150}{\kilo\meter\per\second} $ will also reduce the volumetric bias and thus will tend to reduce the value of $H_0$; however doing so seems unmotivated by the data or theoretical estimates of the PV noise. {We conclude that alternative selection models beyond those discussed are unlikely to significantly reduce our measured value of $H_0$.} 

As the \citetalias{Riess2021ComprehensiveSH0ES} analysis is based on comparison of indicators with much lower relative scatter than the redshifts we use here, the analysis is not likely to be strongly affected by selection effects. Distance selection and biases in the Hubble sample is accounted for using the BBC methodology \citep{KesslerScolnic2017}. However any analysis which makes use of high-scatter indicators, including nearby redshifts as a measure of the Hubble flow, must carefully evaluate the selection of their sample. This has been done in measurements of $H_0$ from ``bright sirens'' in \citet{Abbott2017GravitationalH0} as well as ``dark sirens'' \citep{2019SoaresDESDarkSiren}. \citet{Boruah2020CosmicConstraints} found a nontrivial effect from the use of a volumetric prior in a measurement of $H_0$ from water megamasers using data from \citet{Reid2019}, and we suggest future works more closely examine the selection of {megamaser, SBF, and Tully-Fisher} distance measurements.

Cosmological uncertainties are unlikely to affect the result. We assume Gaussian behavior in our calculation of the peculiar velocity uncertainties, while the cosmological density field is known to show nonlinear and non-Gaussian behavior, particularly at scales $k>\SI{0.1}{\mega\parsec\per\h}$. However the velocity field is comparatively insensitive to these effects, as the velocity field is a factor of $1/k^2$ less sensitive to small scales as compared to the density. {We also neglect any velocity bias, which is absent on large scales in the absence of non-gravitational forces as we deal with in this work (see e.g.,~\cite{Desjacques_biasreview})}. We thus expect our result to be insensitive to the effects quantified by higher order statistics. Our error estimates also assume $\Lambda$CDM physics, but are only weakly dependent on alternative physics through changes to the power spectrum at small scales, given that the power spectrum at large scales has been constrained by numerous independent experiments. Our estimates of the uncertainties in peculiar velocity reconstructions are optimistic, insofar as they assume that the reconstructed velocity fields are no more inaccurate than the differences between the two. At scales larger than the volume of the Cepheid sample, we are unable to effectively constrain these quantities ourselves. At scales smaller than this, we have mitigated these uncertainties by introducing amplitude parameters that are marginalized over in the final result. {This data and others \citep{Said20, Peterson2021PVCorrections} strongly support the use of these reconstructions as broadly applicable models of the nearby universe.}

{This work provides a measurement of $H_0$ independent of any supernovae sample. At the expense of increased uncertainties, dropping supernova measurements from the analysis has three main advantages. First, this work provides a satisfactory systematic error check of the SH0ES analysis: our result is fully compatible with the findings of~\citetalias{Riess2021ComprehensiveSH0ES}, potentially ruling out any sizable systematic error contribution from the inclusion of the supernovae. Further, our measurement of $H_0$ is uncorrelated with the absolute magnitude of higher-redshift supernovae, which circumvents any potential issues when using } this value as a prior in analyses which additionally use Type Ia supernovae as a constraint on the evolution of the distance-redshift relation. Lastly as the sample's maximum redshift is $z \sim 0.01$, our result is robust to any late time effects such as transitions in the dark content of the Universe unless they have occurred within the last $\sim 100$ Myr.

Measurements of the Hubble constant have been conducted at low redshift using a variety of distance indicators. The treatment of peculiar velocities has in some cases been a matter for discussion in the literature \citep{Howlett2020StandardH0,Boruah2020CosmicConstraints,Mukherjee2021VelocitySirens}. Our methodology shares several elements with \citet{Boruah2020CosmicConstraints}, with improvements to the treatment of off-diagonal velocity uncertainties, and a more flexible model framework parametrized to reduce cosmology dependence. We anticipate the use of similar methodology to that presented here in future Cepheid-based measurements of the Hubble constant from the SH0ES team, and suggest these elements should be incorporated into other analyses based on low-redshift distance indicators.

\begin{acknowledgments}

We are thankful to Tamara Davis and Anthony Carr for their contributions to the quality of the Pantheon+ redshifts, and comments on this paper.

Support for this work was provided by NASA through the NASA Hubble Fellowship grant HF2-51462.001 awarded by the Space Telescope Science Institute, which is operated by the Association of Universities for Research in Astronomy, Inc., for NASA, under contract NAS5-26555. This research was supported by NASA/HST 
grant 16269 from the Space Telescope Science Institute. D.S. is supported by DOE grant DE-SC0010007 and the David and Lucile Packard Foundation. This project/publication was made possible through the support of grant \#62314 from the John Templeton Foundation. J.L.B. is supported by the Allan C.\ and Dorothy H. Davis Fellowship.

\end{acknowledgments}

\software{ AstroPy \citep{astropy2013,astropy2018}, corner.py \citep{ForemanMackey2021Corner}, 
Matplotlib \citep{Hunter2007matplotlib}, nbodykit \citep{Hand2018nbodykit}, NumPy \citep{harris2020array}, pystan \citep{pystan}, SciPy \citep{2020SciPy-NMeth}, Stan \citep{Carpenter2017Stan}, tqdm \citep{costa_luis_2021_tqdm}
}
\bibliography{references}{}
\bibliographystyle{aasjournal}

\appendix

\section{Cepheid Methodology Systematics}
\label{app:cephsys}
In order to quantify the systematics associated with methodological choices made internal to the Cepheid analysis, we include additional systematic uncertainties from the analysis variants of \citetalias{Riess2021ComprehensiveSH0ES}. {The baseline SH0ES analysis quantified these effects by measuring $H_0$ for each variant, taking the standard deviation of these as a systematic uncertainty in quadrature. This procedure in general will overestimate the uncertainty, as the analysis is not given the opportunity to mitigate the effect on the measurement by more strongly weighting the data less affected. Since the \citetalias{Riess2021ComprehensiveSH0ES} analysis is generally dependent only on changes in the mean of the calibrator Cepheids relative to the anchor Cepheids, this overestimate would not significantly inflate the final error budget. However our measurement is more sensitive to changes in the relative values of the Cepheid host galaxies due to the correlations among the host galaxy peculiar velocities. By incorporating information about the effect of the analysis variants directly into the model, we minimize the effect on our measurement .}

For a specific effect (i.e. treatment of outliers in the Cepheid P-L relation) we define a set of analysis variants which represent reasonable alternative treatments of the baseline SH0ES Cepheid sample. For an effect {X} represented by $n$ analysis variants, where the $i$th variant produces distance moduli $\boldsymbol{\mu}_\mathrm{X}^{(i)}$, we define a systematic covariance matrix {by the average of the outer products of the residuals relative to the baseline treatment} 

\begin{equation}
    \boldsymbol{\Sigma}^{(\mathrm{X})}_\mu= \frac{1}{n}\sum_{i=0}^n (\boldsymbol{\mu}_\mathrm{X}^{(i)}- \boldsymbol{\mu }_\textrm{Ceph})  \otimes (\boldsymbol{\mu}_\mathrm{X}^{(i)}- \boldsymbol{\mu }_\textrm{Ceph}).
\end{equation}
\noindent The individual covariance matrices are added together to evaluate the total uncertainty in the distance moduli
\begin{equation}
    \boldsymbol{\Sigma}_\mu =  \boldsymbol{\Sigma}^{(\mathrm{SH0ES})}_\mu + \sum_X  \boldsymbol{\Sigma}^{(\mathrm{X})}_\mu.
\end{equation}

We briefly discuss the analysis variants we use here; for full details of the treatment of the data under each, refer to Section 6 and Table 5 of \citetalias{Riess2021ComprehensiveSH0ES}. We budget for the potential systematic impact of four different steps in the Cepheid analysis methodology:

\begin{enumerate}
    \item \textbf{Cepheid Outlier Treatment}: The selection of our Cepheid samples from optical data is expected to leave some residual contamination by blended variable stars in the sample. Our baseline analysis therefore includes outlier rejection by individually discarding the largest outlier until no Cepheids deviate from the global period-luminosity (P-L) fit by $>3.3\sigma$. We include variants 2-8 as viable alternative treatments, excluding variant 9 which discards outlier rejection entirely.
    \item \textbf{Reddening, Extinction, and Color Relations}: Our baseline analysis uses dereddend Wesenheit magnitudes based on the \citet{Fitzpatrick99} reddening law with the absolute-to-selective extinction ratio $R_V$ fixed to 3.3. Alternative treatments include allowing the assumed value of $R_V$ as a free parameter, alternative reddening laws, and individually inferred values of $R_V$ for each Cepheid host galaxy using the empirical dust attenuation framework of \citet{Hahn2022EmpiricalDust}.  We include variants 16,17, 19, 20, 21, and 22 as viable alternative treatments. Variant 23 excludes reddening standardization entirely and is omitted from our budget.
    \item \textbf{Period Standardization}: The baseline sample includes all Cepheids with observed periods $P>\SI{5}{\days}$, and uses a linear P-L relation with a single slope. As some studies have found broken P-L relations for Cepheids at $P=\SI{10}{\days}$, we include a variant (number 24) which allows different slopes above and below $P=\SI{10}{\days}$. The difference in slope is consistent with 0 within $2\sigma$. Other variants exclude high or low period Cepheids from the sample (variants 25 and 26).  All three variants are included as viable alternative treatments.
    \item \textbf{Metallicity}: The baseline variant includes a linear standardization of Cepheid magnitudes for the metallicity of the host galaxy. The free parameter $\gamma$, the slope of the metallicity-luminosity relation, is significant at $\sim 4.6\sigma$. The Cepheid metallicities are inferred based on \citet{Teimoorinia2021MetallicityMachine}. We include a single variant analysis which uses instead the metallicity scale of \cite{Pettini2004AbundanceIndicator}, as discussed further in Section 6.6 and Appendix C of \citetalias{Riess2021ComprehensiveSH0ES}. We exclude variant 30 which discards metallization standardization entirely.
\end{enumerate}

\begin{figure}
    \centering
    \includegraphics{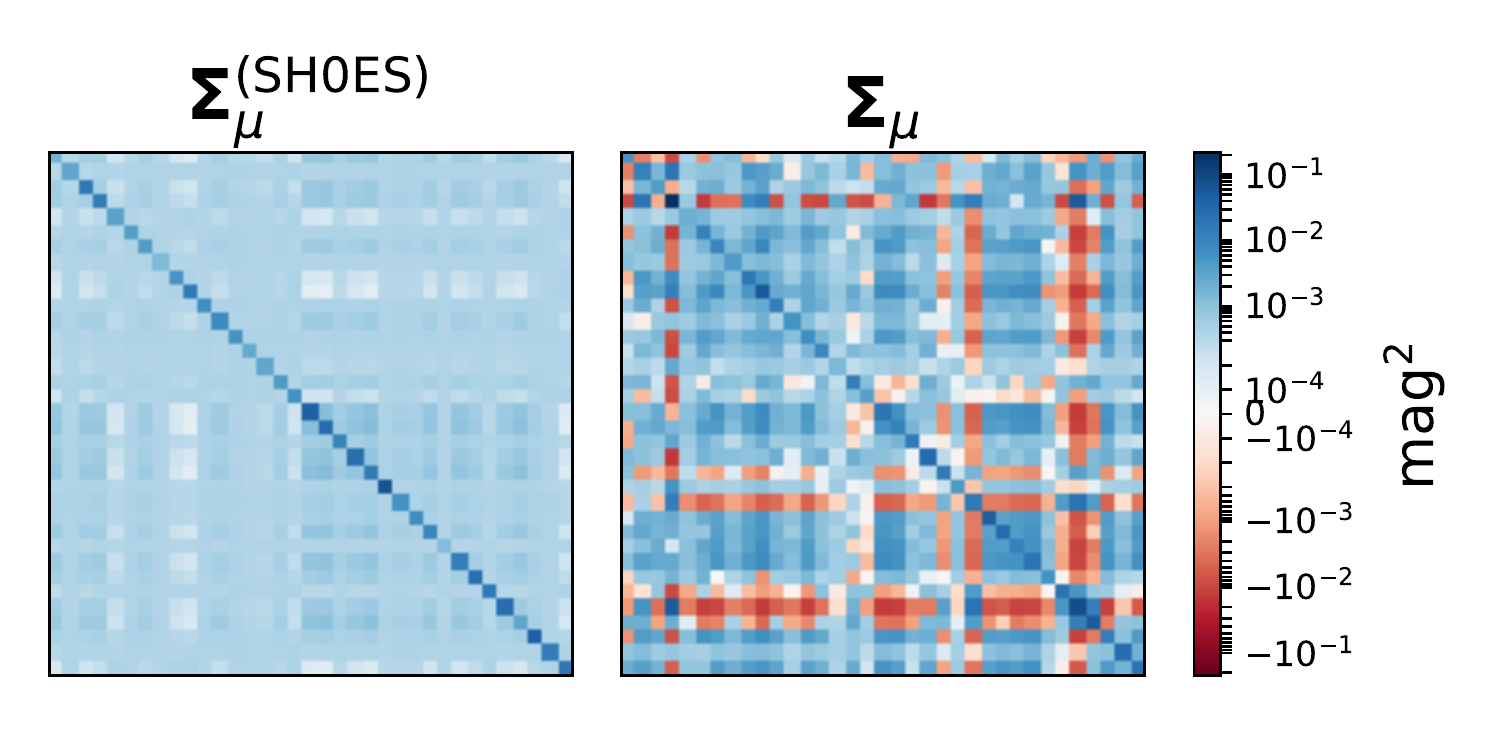}
    \caption{Comparison between the baseline covariance $\boldsymbol{\Sigma}^{(\mathrm{SH0ES})}_\mu$ and the full systematics matrix $\boldsymbol{\Sigma}_\mu$. Galaxies are sorted by distance, with the nearest at the top left. The baseline covariance is visually dominated by the covariance due to the limited number of geometric anchors. Inclusion of the additional systematics internal to the Cepheid sample give the covariance matrix a more complex structure. }
    \label{fig:cephcovcomparison}
\end{figure}

In Figure \ref{fig:cephcovcomparison}, we show both the baseline covariance $\boldsymbol{\Sigma}^{(\mathrm{SH0ES})}_\mu$ and the full  covariance including $\boldsymbol{\Sigma}_\mu$. We show how including each effect discussed above in the systematic covariance matrix affects the mean and variance of the $H_0$ measurement in Table \ref{tab:sysdiffs}. The systematic effect with the most impact on the final measurement is the behavior of the Cepheid period-luminosity relation.

\begin{table}[]
    \centering
    \begin{tabular}{lrrl}
\hline
 Systematics                                       &   $\Delta <H_0>$ from baseline &   $\sigma H_0$ & Contribution to Error Budget   \\
\hline
 $\boldsymbol{\Sigma}^{(\mathrm{SH0ES})}_\mu$ only &                           -0.1 &            2.4 & Ref.                           \\
 Outlier Treatment                                 &                           -0.1 &            2.4 & 0.3                            \\
 Reddening and Color                               &                           -0.2 &            2.4 & 0                           \\
 Period-Luminosity Relation                        &                            0.2 &            2.4 & 0.2                            \\
 Metallicity Scale                                 &                           -0.1 &            2.4 & 0.1                            \\
 $\boldsymbol{\Sigma}_\mu$ (All Sys. Inc.)         &                            0   &            2.5 & 0.7                            \\
\hline
\end{tabular}
    \caption{Mean and standard deviation of $H_0$ as measured with the baseline Cepheid covariance $\boldsymbol{\Sigma}^{(\mathrm{SH0ES})}_\mu$, and with each component of the systematic uncertainty added individually. Error budget contribution defined as $\sqrt{\sigma^2_\mathrm{Effect} - \sigma^2_\mathrm{Ref.}}$. All results used the model variation assuming redshift-limited selection, fixed the peculiar velocity amplitude, and used the \citetalias{Carrick2015} reconstruction, to ensure the assumed errors in the data are kept constant apart from the changes in the Cepheid covariance.
}
    \label{tab:sysdiffs}
\end{table}

\citetalias{Riess2021ComprehensiveSH0ES} presented several other analysis variants. Many of these are not treated here as they impact only the SN\,Ia sample, rather than the Cepheid distances. We also do not make use of analysis variants which ignore large amounts of data (the optical Wesenheit variants, which wholly discards all NIR data) or ignore components of the standardization (such as removing metallicity dependence from the Cepheid luminosity relations). As we are interested in making the most accurate measurement possible given all available data, these analysis variants do not represent reasonable treatments.  In general our final results will  be more sensitive than the \citetalias{Riess2021ComprehensiveSH0ES} result to the exclusion of large amounts of data or the removal of components of the standardization.

\section{Estimation of Covariance in PV-corrected Redshifts}

\label{app:pvcovariancederiv}
\begin{figure}
    \centering
    \includegraphics{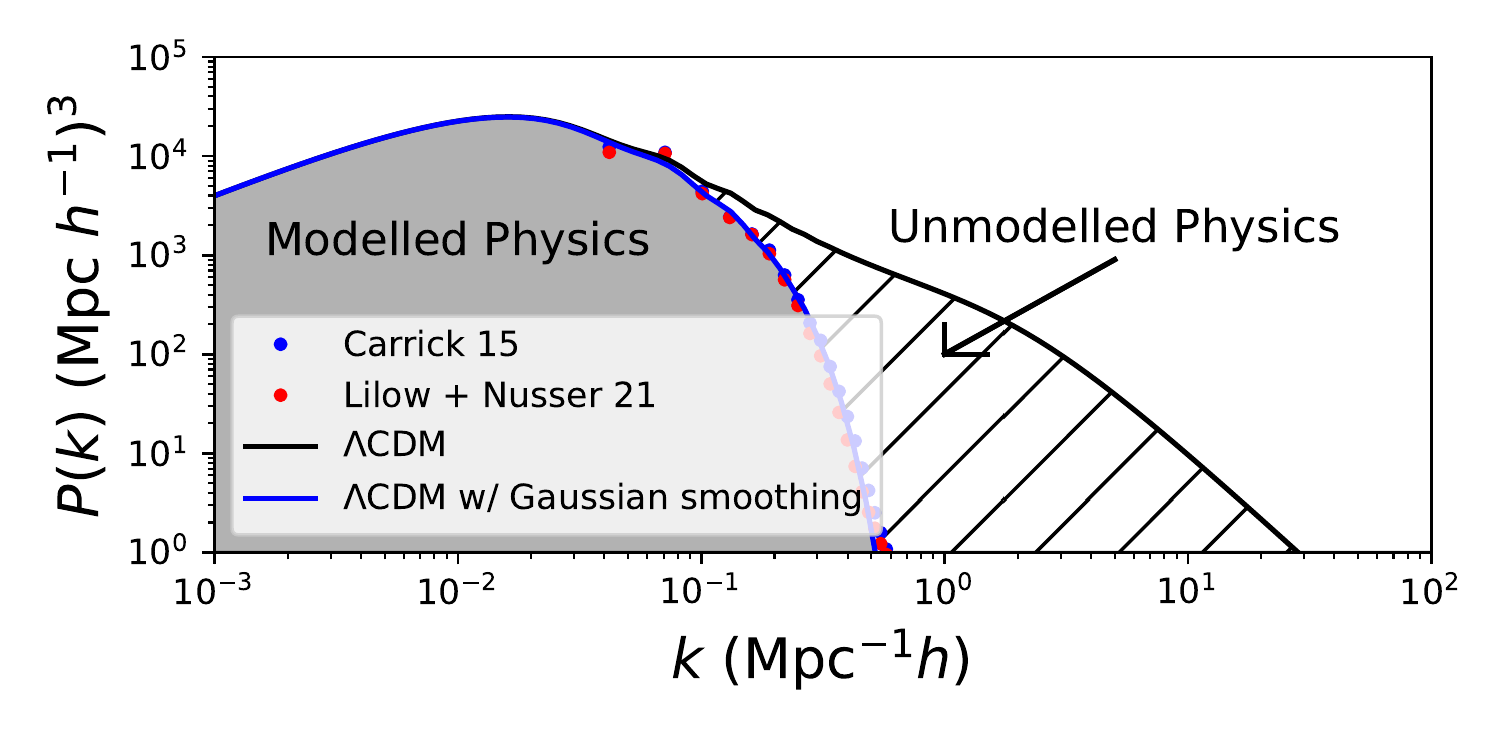}
    \caption{Comparison between the power spectrums of \citetalias{Carrick2015} and \citetalias{LilowNusser2021ConstrainedFlow} and the $\Lambda$CDM spectrum evaluated with CLASS. As would naively be expected, both are well described by a $\Lambda$CDM spectrum smoothed by a Gaussian kernel. The physics represented by the nonlinear part of the power spectrum is smoothed out of the reconstructions, and is thus unmodeled. We take this as a systematic.}
    \label{fig:powerreconcomp}
    \centering
\end{figure}

We first evaluate the power spectrum of the \citetalias{Carrick2015} reconstruction. As the C15 reconstruction has masked some parts of their volume due to a lack of data, we use the \SI{212}{\mega\parsec\per\h} cube centered on the Milky Way. This volume avoids any masking, and encompasses the whole of our sample. We compare with the predicted $\Lambda$CDM power in Figure \ref{fig:powerreconcomp}. As can be seen, the reconstruction's power spectrum is well approximated by $\Lambda$CDM smoothed with a Gaussian kernel at a scale of \SI{7}{\mega\parsec\per\h}. The dispersion of the line-of-sight peculiar velocities within the reconstruction is smaller than the $(\SI{380}{\kilo\meter\per\second})^2$ expected from $\Lambda$CDM: $<v_i^2>\sim (\SI{270}{\kilo\meter\per\second})^2$ evaluated using the corresponding power spectrum. It is clear that the reconstruction cannot correct for the half of the dispersion  that has not been modelled. 
More formally, our quantity of interest are the differences between the estimated and true density/velocity fields, as represented by the density residuals

\begin{equation}
    \delta_{\Delta}(\Vec{r}) = \delta_\mathrm{C15}(\Vec{r}) - \delta_\mathrm{True}(\Vec{r}).
\end{equation}
\noindent Our goal is to determine the power spectrum of this quantity, which will enable us to budget for unmodeled PV noise and quantify the improvement in our result from the use of the reconstructions.  
The power spectrum of the residuals is
\begin{equation}
    P_{\Delta}(k)= P_\mathrm{True}(k) + P_\mathrm{ C15}(k) - 2 P_\mathrm{True , C15}(k),
\end{equation}
where $P_\mathrm{True , C15}$ is the cross-spectrum term

\begin{equation}
     P_\mathrm{True , C15}(k) = \int d^3k^* /(2\pi)^3 <\Tilde{\delta}^*_\mathrm{C15}(\Vec{k}^*) \Tilde{\delta}_\mathrm{True}(\Vec{k}) + \Tilde{\delta}_\mathrm{C15}(\Vec{k}) \Tilde{\delta}^*_\mathrm{True}(\Vec{k}^*) > /2.
\end{equation}
\noindent This term is essentially a measure of the alignment in \textit{phase} between the true and reconstructed density fields, while the power spectrum is phase-independent. In the case where the reconstruction is perfectly correct, this cross spectrum term is equal to the power spectrum, and as we expect the power in the residuals drops to zero. While we can assume an underlying $\Lambda$CDM power spectrum for the true density fields and the power spectrum of the reconstruction can be measured, evaluating the cross-spectral term requires us to determine the ``correctness'' of the reconstructions, which cannot be done from first principles.

As an estimate of this term from the data we have available, we have chosen to base our error budget on the \textit{the differences between} the results of \citetalias{Carrick2015} and \citetalias{LilowNusser2021ConstrainedFlow}. This is likely to give an underestimate of the true uncertainties, as both are based on the 2MRS redshifts published in \citet{Huchra20122MASSRelease}. However this provides an estimate of the methodological uncertainties that includes off-diagonal terms of the covariance, not otherwise available to us. We test this method on our data in Section \ref{sec:results}, as well as parametrizing the scale of covariance in the data and marginalize over these quantities. Ultimately the Cepheid data significantly favor the use of reconstructions with our budgeted uncertainties, and we conclude we have found a reasonable approximation of our true uncertainties. 

By assuming $ P_\mathrm{True , C15}(k) = P_\mathrm{C15 , LN21}(k) $, and using the relation

\begin{equation}
    2 P_\mathrm{C15 , LN21}(k) = P_\mathrm{C15}(k) -  P_\mathrm{C15 - LN21}(k)  + P_\mathrm{LN21}(k), 
\end{equation}
we derive an estimate of the power spectrum which describes the uncertainty in the reconstructed density fields

\begin{align}
    \hat{P}_{\Delta}(k) &=  P_\mathrm{Unmodelled}(k)+ P_\mathrm{Structure}(k) \label{eq:powerdeltaest}\\
    P_\mathrm{Unmodelled}(k)&=P_{ \Lambda\mathrm{CDM}}(k) - P_\mathrm{Reconstructed}(k)\\
    P_\mathrm{Structure}(k) &= P_\mathrm{C15 - LN21}(k)
\end{align}

\noindent where $P_\mathrm{C15 - LN21}(k)$ is the power spectrum of the residuals between the two reconstructions $\delta_\mathrm{C15}(\Vec{r}) - \delta_\mathrm{LN21}(\Vec{r})$. Rather than directly using the power spectrum evaluated from the data cubes, for both \citetalias{Carrick2015} and \citetalias{LilowNusser2021ConstrainedFlow} we use the smoothed $\Lambda$CDM power spectrum
\begin{equation}
    P_\mathrm{Reconstructed}(k)= P_{ \Lambda\mathrm{CDM}}(k) \exp\left( \frac{ -k^2 }{2 \sigma_\mathrm{Smooth}^2 }\right)
\end{equation}
with a smoothing scale $\sigma_\mathrm{Smooth} = \SI{7}{\mega\parsec\per\h}$. Our estimate of the uncertainties then has a fairly simple and intuitive form; subtract the modelled physical information  from the theoretical prediction and add the power describing the differences between our two models. The relative contribution of the two can be seen in Figure \ref{fig:powerestimated}; the power from the differences adds $(\SI{66}{\kilo\meter\per\second} )^2$ of variance, while the nonlinear term contributes $(\SI{266}{\kilo\meter\per\second} )^2$ at much smaller scales. As we have no strong evidence to favor one reconstruction over the other, our error estimate is the same for both, and we produce results using both.

\end{document}